\documentclass[twocolumn,aps,prl,10pt,showpacs,dvips]{revtex4}
\usepackage{epsfig}
\begin{document}
\title{{\Large Microscopic theory of phase transitions in a critical region}}

\author{Vitaly V. Kocharovsky$^{1,2}$ and Vladimir V. Kocharovsky$^{2,3}$}

\affiliation{$^{1}$Department of Physics and Astronomy, Texas A\&M University, College Station, TX 77843-4242, USA\\
$^{2}$Institute of Applied Physics, Russian Academy of Science,
603950 Nizhny Novgorod, Russia\\
$^{3}$Lobachevsky State University of Nizhny Novgorod, 23 Gagarin Avenue, Nizhny Novgorod 603950, Russia}

\date{\today}

\begin{abstract}
    The problem of finding a microscopic theory of phase transitions across a critical point is a central unsolved problem in theoretical physics. We find a general solution to that problem and present it here for the cases of Bose-Einstein condensation in an interacting gas and ferromagnetism in a lattice of spins, interacting via a Heisenberg or Ising Hamiltonian. For Bose-Einstein condensation, we present the exact, valid for the entire critical region, equations for the Green's functions and order parameter, that is a critical-region extension of the Beliaev-Popov and Gross-Pitaevskii equations. For the magnetic phase transition, we find an exact theory in terms of constrained bosons in a lattice and obtain similar equations for the Green's functions and order parameter. In particular, we outline an exact solution for the three-dimensional Ising model.

Published in Physica Scripta {\bf 90}, 108002 (2015), doi:10.1088/0031-8949/90/10/108002.
    
\end{abstract}
\maketitle

\section{1. The central unsolved problem in theoretical physics}

   The problem is to find a microscopic theory of the second order phase transitions that would allow one to follow a formation of an ordered phase from a disordered phase across an entire critical region continuously. The microscopic theory should be derived for a given microscopic Hamiltonian of a mesoscopic or macroscopic system from the first principles of statistical physics and describe an order parameter, correlations as well as other statistical and thermodynamic quantities. Since Landau's breakthrough paper of 1937 \cite{Landau1937,LLV,LL}, it is known that the phase transitions, leading to ferromagnetism, Bose-Einstein condensation (BEC), superfluidity, superconductivity, liquid crystal phases, ferroelectricity and so on, originate from restructuring of a many-body system due to spontaneous symmetry breaking. Landau gave a general mean-field theory of spontaneous symmetry breaking. 

    However, an exact solution for a particular Ising model of a magnetic phase transition in a 2-dimensional (2D) lattice of spins, found by Onsager \cite{Onsager} in 1944 and presented in detail by Yang \cite{Yang} in 1952, showed that the Landau theory was incorrect in the most interesting region surrounding a critical point of a phase transition. According to the Ginzburg-Levanyuk criterion \cite{Ginzburg1960,Levanyuk1959}, found in 1960, in the critical region of a system's parameters, the fluctuations of the order parameter are anomalously large, that is they exceed the mean value of the order parameter. Thus, it became clear that critical phenomena are intimately related to the anomalously large fluctuations and could be described only by a much more involved theory. Such a theory should go far beyond the simplified picture of an averaged, mean field. 
    
    After the development of powerful quantum-field-theory methods in many-body physics in 1950s and 1960s, that problem was attacked by the best theorists both from the point of view of a general theory and the point of view of various particular phase transitions (see, for example, \cite{AGD,HohenbergMartin,Kondor1974,FetterWalecka,Anderson1984,Shi,Shlyapnikov1998,PitString,RevModPhys2004,ProukakisTutorial2008,LLV,LL,PatPokr,Fisher1986,CritPhen-RG1992,Goldenfeld,Kadanoff,Cardy1996,Vicari2002,Berges2002} and references therein). They used all the known machinery of the many-body theory, including Green's functions, Feynman's and Matsubara's diagram techniques, Wick's theorem, Dyson's equation, renormalization group, and so on. In particular, an important understanding of the universality of phase transitions in the vicinity of the critical point came from the renormalization-group approach \cite{Wilson,PatPokr,Fisher1986,CritPhen-RG1992,Goldenfeld,Kadanoff,Cardy1996,Vicari2002,Berges2002}. However, the latter mostly deals with the thermodynamic-limit quantities and phenomenological scaling and universality, does not fully account for the constraints of a many-body Hilbert space (in particular, due to an employment of the grand-canonical-ensemble or similar approximations), and focuses only on a few terms of an intermediate asymptotic expansion at the wings of the critical region, related mainly to the critical exponents. The renormalization-group approach does not provide a full solution that would be valid both inside the entire critical region and outside it, in the disordered as well as ordered phases. 

    Thus, the current theory of critical phenomena has a well-established microscopic basis and is complete conceptually, but there is the necessity to develop new, regular, and more powerful methods for finding the full solution. The most striking fact is an absence of a solution to the major problem of how the systems go through a critical point. In general, it is not known for both equilibrium and nonequilibrium phase transitions. Here the principle difficulty lies in an invalidity of the usually used equations for the order parameter, correlation functions and other related quantities in the central part of the critical region. For example, for the BEC neither the state of the art equations for the order parameter and the Green's functions (the Gross-Pitaevskii and Beliaev-Popov equations) for the equilibrium case \cite{Shi,Shlyapnikov1998,PitString,RevModPhys2004,ProukakisTutorial2008}, nor the quantum kinetic equations for the nonequilibrium case (see the most advanced analysis in \cite{Reichl}) are valid in the critical region. 

    Here we should comment on the terminology, adopted in the literature (see, e.g., \cite{PatPokr,Kadanoff}) and employed in the present paper, where the renormalization group approach is named phenomenological to underline its concept of the effective Hamiltonians and effective fields. The latter describe only the long-range correlations in the system at the cost of a rescaling of the short-range fluctuations. In fact, that rescaling (i.e., averaging or course-graining) procedure constitutes a subtle, often not fully justified and mathematically controlled, approximation in that theory. The exact approach, which takes care of the fields and fluctuations at all scales, is named microscopic. Of course, both approaches constitute the microscopic theory of phase transitions and are based on the same basic principles of the quantum or classical statistical physics of many-body systems, governed by some microscopic interaction Hamiltonian. That terminology is not perfect, but it catches the fact that the renormalization group starts with an approximation of the system and then attempts to find the solution, which could represent only some intermediate asymptotics of the exact full solution. On the contrary, the exact approach starts with the rigorous equations, relations and solutions for a mesoscopic, finite system and then analyzes them. It yields, in principle, more detailed and full information on the critical phenomena and critical functions. In particular, it goes beyond the thermodynamic-limit critical exponents or intermediate power-law asymptotics of the critical-region wings. 

    Even the exact, nonperturbative renormalization group equations (for a review, see \cite{Vicari2002,Berges2002}), including the ones derived by Wilson \cite{Wilson} via an $\varepsilon$-expansion of a space dimensionality $d=4-\varepsilon$ for a $s^4$-model, are aimed to approximate the original microscopic Hamiltonian by an effective Hamiltonian for a large-scale part of a field. Moreover, these exact equations are very difficult to implement due to their complexity. When Wilson tried to show "that the exact renormalization group equations are not hopelessly intractable functional equations" \cite{Wilson}, he still substituted them with an approximate recursion formula which is mathematically uncontrollable, but yields the first few orders of the $\varepsilon$-expansion. So, one must perform approximations and/or truncations, such as in a well-developed method of an effective average action \cite{Berges2002}. In fact, a number of approximations for the renormalization are introduced, including a neglect method, decimation and other real-space renormalization schemes as well as field-theoretical techniques. However, their rigorous relation to a full exact solution of the mesoscopic problem is still not completely understood.

    Simultaneously, tremendous effort was spent to study the problem of phase transitions experimentally and numerically in a countless number of various particular many-body systems, including macroscopic and mesoscopic systems, condensed matter, trapped gases, cosmology, relativistic fields, high energy particles, etc. There is hardly any other problem in physics that attracted such an enormous amount of interest and effort. Suffice it to say that more than thirty Nobel prize winners received their Nobel awards for the studies related to phase transitions and spontaneous symmetry breaking, including Onnes (1913), Landau (1962), Feynman (1965), Onsager (1968), N$\acute{e}$el (1970), Bardeen (1972), Cooper (1972), Schrieffer (1972), Josephson (1973), Anderson (1977), Mott (1977), Prigogine (1977), Kapitza (1978), Weinberg (1979), Glashow (1979), Salam (1979), Wilson (1982), Bednorz (1987), M$\ddot{u}$ller (1987), de Gennes (1991), Lee (1996), Osheroff (1996), Richardson (1996), Cornell (2001), Ketterle (2001), Weiman  (2001), Abrikosov (2003), Ginzburg (2003), Leggett (2003), Nambu (2008), Kobayashi(2008), Maskawa (2008), Englert (2013), Higgs (2013). Hundreds of books and reviews as well as many thousands papers have been devoted to this field of research.
    
    Why is the problem of the critical region so important for the theoretical physics? An answer follows from a fact that in all other regions of parameters the macroscopic systems show a routine behavior with the normal, Gaussian thermodynamic fluctuations, described either by classical or quantum statistical mechanics. The former was understood in the beginning of the twentieth century due to the works by Boltzmann, Gibbs et al. The latter was understood by 1950s after the development of quantum mechanics (by 1930s) and quantum field theory (by 1950s). Many-body systems can acquire novel nontrivial properties only in these critical regions, where the microscopic interaction between particles results in a system's self-organization into an ordered phase. That process restructures the system's configuration and breaks a symmetry, generally, via an internal instability. Thus, a solution to the problem of finding a microscopic theory of phase transitions in the critical region became the major key to progress in our understanding of the macroscopic and mesoscopic systems. 

    All real systems in the experiments and applications, especially in the modern nanotechnologies, are finite. Their critical regions are also finite and constitute, in fact, the most interesting and important regions of parameters. Resolving the structure of the critical region is equivalent to solving the problem for a mesoscopic (large, but finite) system. The critical region cannot be treated as just a discontinuity point, as it was assumed in a standard, bulk-limit analysis of the thermodynamic fluctuations and the proofs of an equivalence of different Gibbs ensembles (see \cite{Khinchin1943,Ruelle1969,Zubarev1971,Dobrushin1973,Lebowitz1978,Martin-Lof1979,EllisJStatPhys2000} and references therein).
    
    A fact, that the singularities are rounded off in the mesoscopic systems, is well known since the Ehrenfest's criticism of the Einstein's theory of a phase transition in a gas of Bose particles and is relevant to the numerous, especially modern experiments and simulations. Its proper description requires the theoretical, computational tools beyond a thermodynamic-limit analysis. Moreover, the solution of the mesoscopic problem is needed for the computation of the thermodynamic-limit critical functions in a close vicinity of the critical point. The latter functions describe a continuous transition through the critical point and have a self-similar scaling, different from the power-law scaling at the wings of the critical region. The rigorous results of the presented in \cite{PLA2015} and in this paper microscopic theory of critical phenomena provide such computational tools and complement the renormalization-group finite-size scaling approach. 

    Moreover, the solution of the problem for the equilibrium systems, which is a subject of the present paper, is crucially important also for the theory of more complicated and diverse nonequilibrium phase transitions. This theory should describe the formation of an ordered phase in various nonequilibrium processes and is even less studied than the equilibrium one (see, e.g., \cite{Zubarev1971,Zubarev1996}).  
    
    Unfortunately, despite enormous effort, interest, and importance, the formulated above problem of finding a full microscopic theory of phase transitions has not been solved. Such a theory, that should start from a microscopic Hamiltonian and derive the closed universal equations capable to provide the full solution connecting the asymptotics of the ordered and disordered phases across the entire critical region, has not been found so far even for any one of the numerous types of phase transitions.

\section{2. How to get a correct theory of the critical phenomena}

\noindent Recently we found a general solution to this problem \cite{PLA2015}. In the present paper, we formulate that microscopic theory for two cases: Bose-Einstein condensation in an interacting gas (Sect. 3) and ferromagnetism in a lattice of spins, interacting via a Heisenberg or Ising Hamiltonian (Sect. 4 - 6). We derive the fundamental equations for the order parameter and Green's functions, which are valid not only in the fully ordered, low-temperature phase, but also in the entire critical region and in the disordered phase. These equations allow one to resolve a fine structure of the system's statistics and thermodynamics near a critical point both for the mesoscopic and macroscopic systems. Also, they allow one to find the asymptotics at the wings of the critical region and, in particular, the critical exponents and the critical functions. 

    Except for some degenerate cases, like a BEC in an ideal gas, a generic process of symmetry breaking is related to an instability in a many-body system that originates from a microscopic interaction between individual particles. The instability starts from the spontaneous or thermal fluctuations, leads to a growth of the order parameter to a macroscopic level, and ends with a formation of an ordered equilibrium phase with a broken symmetry due to a nonlinear saturation of the instability. The complex nature of critical phenomena in phase transitions is not reduced even to an interplay of the many-body, instability and nonlinear phenomena. Those processes are subject to a whole set of the simultaneously present and equally important factors, which should be properly taken into account in a correct microscopic theory. 
     
    First of all, there are the constraints of a many-body Hilbert space. These fundamental constraints could originate from some local properties of the individual particles or from some nonlocal, global properties of a system's configuration. (Do not mix the term "constraints of the Hilbert space" with the external parameters or fields in a Hamiltonian, like a trap potential $U_{ext}$ or an external magnetic field $B_{ext}$, which sometimes are also referred to as constraints in the literature \cite{CritPhen-RG1992}.) Their relation with the integrals of motion, prescribed by a broken symmetry in virtue of a Noether's theorem, is also important. 

    An example of local constraints is given by the occupation numbers $n_{\bf r}$ of the spin bosons, constituting the spins in the lattice of a ferromagnet. According to Holstein and Primakoff \cite{HolsteinPrimakoff}, they are constrained to an interval of values $0 \leq n_{\bf r} \leq 2s$, where $s$ is the spin value. 

    An example of global constraints is a particle-number constraint for a system of $N$ particles in a trap. That constraint is responsible, via a related constraint-cutoff mechanism, for the very existence of the BEC phase transition and its nonanalytical features \cite{Anderson1984,PRA2010,JPhys2010,PRA2014,JPhys2014}. The latter implies an insufficiency of a grand-canonical-ensemble approximation, which is incorrect in the critical region \cite{HohenbergMartin,ProukakisTutorial2008} because of averaging over the systems with different numbers of particles, both below and above the critical point. That averaging over the condensed and noncondensed phases at the same time implies a relative error on the order of unity for the critical functions. This is well known, but it is very hard technically to go on with the canonical ensemble. 

\begin{figure}[h]
\includegraphics[width=82mm]{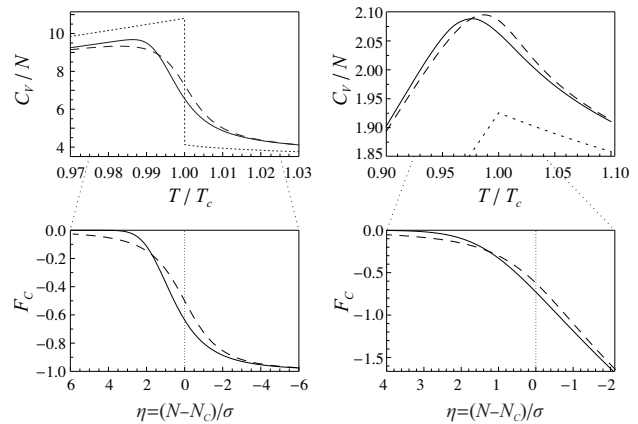}
\caption{A failure of the grand-canonical-ensemble approximation (dashed lines) to yield an exact canonical-ensemble specific heat (solid lines) for an ideal gas of $N=10^4$ particles, confined in the 3D box with zero Dirichlet boundary conditions (right) and in the isotropic harmonic 3D trap (left). The upper plots show the specific heat $c_V=C_V/N$ versus a deviation of temperature from its critical value $T_c$. The lower plots show a critical function $F_C(\eta) = a [c_V -c_V(\mu =0)]$ that is a deviation of the specific heat from its value at zero chemical potential scaled by a suitable factor $a$, making variation of $F_C$ within a critical region on the order of unity. The lower plots yield also a universal structure of the specific heat in the thermodynamic limit, since the critical function $F_C(\eta)$ quickly reaches a limiting form (i.e., stops changing) in that limit due to a use of a self-similar variable $\eta$. A thermodynamic limit of the grand-canonical-ensemble ansatz in the usually used approximation of continuous spectrum is shown on the upper plots by the dotted lines and is even more off the exact result.}
\label{fig:C}
\end{figure}

    Again, the main problem is a computational, theoretical one. Namely, it is to find a rigorous solution, compatible with the particle-number constraint. Moreover, there is a conceptual problem behind the deviation of the grand-canonical-ensemble approximation from the canonical-ensemble result. Namely, a failure of the grand-canonical-ensemble approximation in the critical region persists even in the thermodynamic limit of a macroscopically large system. It is demonstrated in Fig. \ref{fig:C} for the specific heat of an ideal Bose gas in the box and isotropic harmonic traps. The exact canonical-ensemble and grand-canonical-ensemble specific heats, shown in Fig. \ref{fig:C}, are known from an analytical solution for the statistics and thermodynamics of an ideal-gas BEC in arbitrary trap. It was recently found in \cite{PRA2010,JPhys2010,PRA2014,JPhys2014} and is indistinguishable from the numerically calculated curves in the critical region.

     It means that the grand-canonical-ensemble approximation does not allow one to correctly describe a continuous phase transition through the critical point. That fact is not usually appreciated, probably, in view of an absence of any alternative theoretical tools to compute the result within the canonical ensemble. The present theory cures that drawback. 

    Note that, even in the thermodynamic limit, the critical region has an order-of-unity width in terms of a self-similar variable $\eta=(N-N_c)/\sigma$. The latter represents a deviation from the BEC critical point in terms of a difference between the number of trapped particles $N$ and its critical value $N_c$, scaled by a dispersion of fluctuations in the number of condensed particles $\sigma$. That variable resolves a fine universal structure of all thermodynamic and statistical critical functions in the critical region. This structure has a well-defined, smooth limiting form, despite a fact that a size of the critical region in terms of an absolute temperature $T$ is shrinking to zero in the thermodynamic limit.

    An absence of an acceptable diagram technique for the constrained many-body systems (in particular, for the canonical ensemble), which has to be used instead of the standard, unconstrained diagram technique based on the grand canonical ensemble, partly explains why the microscopic theory of phase transitions was not found for such a long time. Despite many attempts \cite{KwokWoo}, the required constrained diagram technique has not previously been found.
     
    Second, one has to solve the problem for a finite system with a mesoscopic (i.e., large, but finite) number of particles $N$. It is necessary in order to correctly calculate an anomalously large contribution of the lowest energy levels to the critical fluctuations and to avoid the infrared divergences of the standard thermodynamic-limit approach \cite{Shi,PitString,RevModPhys2004,ProukakisTutorial2008,LLV,LL,PatPokr} as well as to resolve a fine structure of the critical region. A finite-size scaling technique (see, for example, the reviews \cite{Cardy1996,Vicari2002}) is aimed at this task, but it is phenomenological and mainly oriented to a numerical analysis of the critical scaling.
     
    Last, but not least, is an important fact that in the critical region there are no closed Dyson-type equations for the constrained, true Green's functions. A crucial finding of the present theory is the exact recurrence equations for the partial one- and two-operator contractions. They follow from a remarkable property of the partial operator contractions to reproduce themselves under a contraction operation.
     
    The previously developed methods did not allow one to account simultaneously for all these complex factors. We introduce the required additional techniques, including the following novel methods: 
  
   (i) the nonpolynomial diagram technique and partial contraction of operators \cite{KochLasPhys2007,KochJMO2007,PLA2015}, 

   (ii) the partial difference (recurrence) equations for the operator contractions \cite{PLA2015} (a discrete analog of the partial differential equations in discrete mathematics \cite{PDE-Cheng,DE-Agarwal,DE-Elaydi}),

   (iii) a characteristic function and cumulant analysis \cite{PRA2000,PRA2010} for a joint distribution of the noncommutative observables. 

   They are required by the very nature of the critical correlations in the constrained systems as well as by the structure of the rigorous many-body theory. They complement the standard Matsubara technique of the Green's functions and Dyson equation \cite{LL,AGD,FetterWalecka}. 
   
   The mean field and related approximations, associated with the Landau, Bogoliubov, Beliaev-Popov and similar standard diagrammatic approaches, lead to the incorrect results for the critical behavior and critical exponents. The renormalization group approach to the constraints in the systems with the critical behavior is also approximate, although it allows one to catch the main power-law critical asymptotics at the wings of the critical region for the models like the non-linear sigma model or the Heisenberg model of ferromagnetism.
   
   The main point, however, is that this renormalization group approach is very different from the exact analysis of the theory presented in \cite{PLA2015} and below. In particular, the latter allows us to formulate the theory of critical phenomena in the spin systems in terms of the constrained spin bosons and treat those constraints exactly. In that way, we obtain the exact results, like the ones for the 3D Ising model (Sect. 5), which yield not only the critical exponents, but a much more full description of the critical phenomena in the entire critical region. Moreover,  we can calculate the critical functions and exponents for the 3D systems by obtaining them from those exact solutions, that is, by a regular Green-function's method. It should be compared with the currently used approximate methods, based, for example, on the physically artificial Wilson's $\varepsilon$-perturbation of the space dimensionality $d=4-\varepsilon$, real-space renormalization group, high-temperature series, Borel's summation, or numerical simulations.

    In retrospect, a desire to build a whole theory around just one or two "main ideas", simplifications or approximations, while missing the other similarly important factors of the complex critical phenomenon in phase transitions, was a typical reason for the failure of many previous attempts to solve the problem. No one of the principle factors could be missed, if we have to derive the correct microscopic theory in the critical region for a given microscopic Hamiltonian from the first principles of statistical physics. 
    
    An example of a failure in the critical region and a restriction of an analysis to just a well-formed ordered phase is Dyson's theory of spin waves in a ferromagnet \cite{Dyson1956}. Apparently, due to a lack of the proper mathematical apparatus, first of all, the partial contraction of operators and diagram technique for the nonpolynomial averages, Dyson thought that "the Holstein-Primakoff formalism is thus essentially nonlinear and unamenable to exact calculations". In Sect. 4 we refute Dyson's conclusion and show how to solve the problem. 

    A well-known "Theory of many-body systems: Pair theory" by Arnowitt and Girardeau \cite{GA,Girardeau1998}, that was stimulated by a Bardeen-Cooper-Schrieffer theory of superconductivity, is another example of such a "one-main-idea" approximation.

    Another possible reason is a prejudicial disbelief of many theorists in the very possibility to find a rigorous full solution to such a complicated many-body problem. The latter is clearly seen, for example, in an attitude of Bogoliubov to the problem of finding the microscopic theory of BEC and superfluidity \cite{Bogoliubov1947,BogoliubovLectures1}. A nice exposition on the Bogoliubov approach is given in a review \cite{Zagrebnov}. In most works on the BEC and superfluidity, the rigorous microscopic theory in the critical region is not even addressed due to its complexity (see \cite{AGD,HohenbergMartin,FetterWalecka,Kondor1974,Anderson1984,Shi,Shlyapnikov1998,PitString,RevModPhys2004,ProukakisTutorial2008,BogoliubovLectures1,Zagrebnov} and references therein). For example, Hohenberg and Martin in their famous review on the microscopic theory \cite{HohenbergMartin} explicitly state that "We have nothing to say on the difficult problem of the phase transition".

    A recent trend to substitute a consistent microscopic theory of a phase transition in a given system by an approximate calculation of just a restricted subset of its universal features for a phenomenological model, for example, the critical exponents, also contributed to keeping the problem unsolved.

    The present paper is dedicated to a great memory of our colleague Dick Arnowitt and other pioneers of the many-body theory, who paved a way to our understanding of the many-body physics. The paper is based on the authors' talk presented at the Symposium in Honor of Dr. Richard Arnowitt, Texas A$\&$M University, College Station, Texas, USA, September 19-20, 2014.

\section{3. Bose-Einstein condensation in an interacting gas: The rigorous microscopic theory}
     
    The BEC originates from a conservation of the total number of particles $N$ in a system:
\begin{equation}
\hat{n}_0 + \hat{n} = N; \qquad \hat{n} = \sum_{{\bf k}\neq 0}\hat{n}_{\bf k}.
\label{N}
\end{equation}
Here $\hat{n}_0 = \hat{a}_0^{\dagger }\hat{a}_0$ and $\hat{n}_{\bf k}=\hat{a}_{\bf k}^{\dagger }\hat{a}_{\bf k}$ are the occupation operators of the nondegenerate ground state ${\bf k}=0$ and the excited states $({\bf k}\neq 0)$ in a trap. The creation, $\hat{a}_{\bf k}^{\dagger }$, and annihilation, $\hat{a}_{\bf k}$, operators for a ${\bf k}$-eigenstate in a trap obey a canonical commutation relation $[\hat{a}_{\bf k}, \hat{a}_{\bf k'}^{\dagger }]=\delta_{{\bf k},{\bf k'}}$, where $\delta_{{\bf k},{\bf k'}}$ is a Kronecker's delta. Note that a system of photons in equilibrium inside a black-box cavity does not demonstrate the BEC since it lacks a gauge symmetry due to absorption in the cavity's walls and, hence, the number of photons is not its integral of motion. 
   
    Thus, a rigorous microscopic theory of BEC has to be started with an exact reduction of a many-body Hilbert space. It is restricted by the particle-number constraint, which is an integral of motion in virtue of a global gauge symmetry to be broken in the BEC phase transition. 

\subsection{3.1. The constrained many-body Hilbert space, the true canonical-ensemble excitations, and the Hamiltonian with a constraint interaction}

    Let us consider the BEC of $N$ interacting particles with a mass $M$ in a cubic box with volume $V=L^3$ and periodic boundary conditions. The trap's one-particle eigenfunctions $L^{-3/2}e^{i{\bf kr}}$ and energies $\varepsilon_{\bf k}=\frac{\hbar^2 k^2}{2M}$ are specified by a wave-vector quantum number ${\bf k}=\{ k_x , k_y , k_z \}, k_i = \frac{2\pi}{L}q_i$, where $q_i=0, \pm 1, \dots$ is an integer.

    The reduction of an unconstrained Fock space $\mathcal{H}^{(0)}$, which spans all Fock states $|n_{\bf k}\rangle$ with integer occupations $n_{\bf k}\in [0,\infty )$, to a physical, canonical-ensemble subspace $\mathcal{H}_N^{(0)}$ is achieved in two steps: 
\begin{equation}
\mathcal{H}^{(0)} \rightarrow \mathcal{H}^{(0)}_{{\bf k}\neq 0} \rightarrow \mathcal{H}_N^{(0)}.
\label{reduction}
\end{equation}
First, we restrict the Fock space $\mathcal{H}^{(0)}$ to a Fock space $\mathcal{H}^{(0)}_{{\bf k}\neq 0}$ with an excluded ground-state component, since its occupation $n_0 = N-n$ for any Fock state is determined by the total excited-states occupation $n=\sum_{{\bf k}\neq 0}n_{\bf k}$. Second, we limit an allowed number of excitations by the total number of particles loaded in the trap, $n\leq N$. 

    According to \cite{PLA2015,PRA2010}, the physics (dynamics, fluctuations, etc.) of the system is determined by the creation and annihilation of the canonical-ensemble excitations (not particles!) via the operators 
\begin{equation}
\hat{\alpha}_{\bf k}^{'\dagger}=\theta(N-\hat{n})\hat{\alpha}_{\bf k}^{\dagger}, \qquad \hat{\alpha}^{'}_{\bf k}=\hat{\alpha}_{\bf k}\theta(N-\hat{n}), 
\label{alpha'}
\end{equation}
which leave invariant (i.e., do not lead out of) the physical, canonical-ensemble subspace $\mathcal{H}_N^{(0)}$. They are the step-function $\theta(N-\hat{n})$ cutoff of the transition operators 
\begin{equation}
\hat{\alpha}_{\bf k}^{\dagger}=\hat{a}_{\bf k}^{\dagger}(1+N-\hat{n})^{-1/2}\hat{a}_0, \quad \hat{\alpha}_{\bf k}=\hat{a}_0^{\dagger}(1+N-\hat{n})^{-1/2}\hat{a}_{\bf k}, 
\label{alpha}
\end{equation}
which include a proper normalization; $\theta(x)=1$ if $x\geq 0$ and $\theta(x)=0$ if $x<0$. The operators in Eq. (\ref{alpha}) were introduced in the theory of BEC by Bogoliubov \cite{Bogoliubov1947} in 1947 as well as by Girardeau and Arnowitt \cite{GA,Girardeau1998} in 1959 and coincide with the used earlier, in 1940, by Holstein and Primakoff \cite{HolsteinPrimakoff} operators for the spin bosons, if a value of a spin is equal to $s=N/2$ (see Sect. 4).

    We introduce also a Fock space $\mathcal{H}$, which spans only the excited states $|n_{\bf k\neq 0}\rangle$, and the creation and annihilation operators in that space $\mathcal{H}$, $\hat{\beta}_{\bf k}^{\dagger}$ and $\hat{\beta}_{\bf k}$, which obey a canonical commutation relation $[\hat{\beta}_{\bf k}, \hat{\beta}_{\bf k'}^{\dagger }]=\delta_{{\bf k},{\bf k'}}$. Their $\theta(N-\hat{n})$-cutoff counterparts are the subspace $\mathcal{H}_N$, which spans only the Fock states with a total occupation $n\in [0,N]$ not larger than the number of particles $N$, and the operators
\begin{equation}
\hat{\beta}_{\bf k}^{'\dagger}=\theta(N-\hat{n})\hat{\beta}_{\bf k}^{\dagger}, \qquad \hat{\beta}^{'}_{\bf k}=\hat{\beta}_{\bf k}\theta(N-\hat{n}).
\label{hatbeta'}
\end{equation} 
They constitute a representation, which is exactly isomorphic to the original one in Eq. (\ref{alpha'}) and will be used from now on. In the sense of that isomorphism, one has
\begin{equation}
\hat{\alpha}_{\bf k}^{'}=\hat{\beta}_{\bf k}^{'}, \ \hat{\alpha}_{\bf k}^{'\dagger}=\hat{\beta}_{\bf k}^{'\dagger}, \ \hat{n}_{\bf k}=\hat{\alpha}_{\bf k}^{'\dagger}\hat{\alpha}_{\bf k}^{'} =\hat{\beta}_{\bf k}^{'\dagger}\hat{\beta}_{\bf k}^{'} \ \text{on} \ \mathcal{H}_N^{(0)}=\mathcal{H}_N.
\label{alpha'=beta'}
\end{equation}
The isomorphism in Eq. (\ref{alpha'=beta'}) is not trivial and is valid only on $\mathcal{H}_N^{(0)}=\mathcal{H}_N$. First, the commutation relations for the creation and annihilation operators of the true excitations in Eqs. (\ref{alpha'}) and (\ref{hatbeta'}) are not canonical:
\begin{equation}
[\hat{\alpha}_{\bf k}^{'}, \hat{\alpha}_{\bf k'}^{'\dagger}]=\delta_{{\bf k},{\bf k'}}(1-\delta_{\hat{n},N})-\hat{\alpha}_{\bf k'}^{'\dagger} \hat{\alpha}_{\bf k}^{'} \delta_{\hat{n},N},
\label{nonCCRalpha}
\end{equation}
\begin{equation}
\qquad [\hat{\beta}_{\bf k}^{'}, \hat{\beta}_{\bf k'}^{'\dagger}]=\delta_{{\bf k},{\bf k'}}(1-\delta_{\hat{n},N})-\hat{\beta}_{\bf k'}^{'\dagger} \hat{\beta}_{\bf k}^{'} \delta_{\hat{n},N}.
\label{nonCCR}
\end{equation}
Second, it does not imply an isomorphism between the operators $\hat{\alpha}_{\bf k}, \hat{\alpha}_{\bf k}^{\dagger}$ on $\mathcal{H}^{(0)}_{{\bf k}\neq 0}$ and the operators $\hat{\beta}_{\bf k}, \hat{\beta}_{\bf k}^{\dagger}$ on $\mathcal{H}$, because a commutation relation for the latter is canonical, but a commutation relation for the former coincides with the noncanonical one in Eq. (\ref{nonCCRalpha}).

    A free (zeroth-order), ideal-gas Hamiltonian in the canonical-ensemble Hilbert space remains standard:
\begin{equation}
H_0 = \sum_{{\bf k}\neq 0} \varepsilon_{\bf k} \hat{n}_{\bf k}, \qquad \hat{n}_{\bf k}=\hat{\beta}_{\bf k}^{\dagger}\hat{\beta}_{\bf k}.
\label{H_0}
\end{equation}

    However, the interaction Hamiltonian $H_{N}^{'}$ in the physical, constrained Hilbert subspace $\mathcal{H}_N^{(0)}=\mathcal{H}_N$ acquires an additional, induced by the constraint, nonlinear interaction determined by the terms with the nonpolynomial operator functions $\theta(N-\hat{n})$ and $\sqrt{N-\hat{n}}$. An explicit formula for $H_{N}^{'}$ follows from a standard two-body interaction Hamiltonian \cite{AGD,HohenbergMartin,Kondor1974,FetterWalecka,LLV,LL,Anderson1984,Shi,Shlyapnikov1998,PitString,RevModPhys2004,ProukakisTutorial2008}
\begin{equation}
H^{'} =\int \hat{\Psi}^{\dagger}({\bf r_1}) \hat{\Psi}^{\dagger}({\bf r_2}) U({\bf r_1}-{\bf r_2}) \hat{\Psi}({\bf r_2}) \hat{\Psi}({\bf r_1}) \frac{d^3 r_1 d^3 r_2}{2}.
\label{standardHint}
\end{equation}
The latter is written in the original, unconstrained, auxiliary Fock space $\mathcal{H}^{(0)}$ via a field operator $\hat{\Psi}({\bf r})=\frac{1}{\sqrt{V}}\sum_{{\bf k}} \hat{a}_{\bf k} e^{i{\bf kr}}$ at  a position ${\bf r}$ and assuming a symmetric potential $U({\bf r_1}-{\bf r_2})=U({\bf r_2}-{\bf r_1})$. An exact result is \cite{PLA2015}
\begin{equation}
H_{N}^{'} =\int \hat{\Psi}^{'\dagger}_N({\bf r_1}) \hat{\Psi}^{'\dagger}_{N-1}({\bf r_2}) U \hat{\Psi}^{'}_{N-1}({\bf r_2}) \hat{\Psi}^{'}_N({\bf r_1}) \frac{d^3 r_1 d^3 r_2}{2},
\label{Hint}
\end{equation}
 where we use the true, step-function cutoff field operators
\begin{equation}
\hat{\Psi}^{'}_N({\bf r})=\hat{\Psi}_N({\bf r})\theta(N-\hat{n}), \quad \hat{\Psi}_N({\bf r})=\sqrt{\frac{N-\hat{n}}{V}}+\hat{\beta}_{\bf r}, 
\label{PsiN}
\end{equation}
\begin{equation}
\hat{\beta}_{\bf r}=\frac{1}{\sqrt{V}}\sum_{{\bf k}\neq 0} \hat{\beta}_{\bf k} e^{i{\bf kr}}, \quad \hat{\beta}_{\bf r}^{\dagger}=\frac{1}{\sqrt{V}}\sum_{{\bf k}\neq 0} \hat{\beta}_{\bf k}^{\dagger} e^{-i{\bf kr}}.   
\label{beta_r}
\end{equation}

   For any operator $\hat{A}$, we introduce a usual Matsubara operator $\tilde{A}_{\tau}=e^{\tau H}\hat{A}e^{-\tau H}$, which evolves in the Heisenberg representation in accord with a total Hamiltonian $H=H_0 +H_{N}^{'}$ in an imaginary time $\tau \in [0, \frac{1}{T}]$, related to an inverse temperature $1/T$.
   
   A primary question to a microscopic theory is a derivation of the equations for the mean value and correlations of the Matsubara annihilation and creation operators \cite{AGD,HohenbergMartin,Kondor1974,FetterWalecka,LLV,LL,Anderson1984,Shi,Shlyapnikov1998}, both the unconstrained ($\tilde{\beta}_{\kappa}, \tilde{\bar{\beta}}_{\kappa}$) and true ($\tilde{\beta}^{'}_{\kappa}=\tilde{\beta}_{\kappa}\theta(N-\tilde{n}_{\tau}), \tilde{\bar{\beta}}'_{\kappa} =\theta(N-\tilde{n}_{\tau})\tilde{\bar{\beta}}_{\kappa}$) ones. Here $\kappa=\{\tau,{\bf k}\}$ and the operators are written in a momentum representation. Below we use also a coordinate representation, where $\kappa$ is replaced by $x=\{\tau,{\bf r}\}$. The definitions of the unconstrained (auxiliary) and true coherent order parameters and Matsubara Green's functions are, respectively, as follows
\begin{equation}
\bar{\beta}_{\bf k}=\langle \hat{\beta}_{\bf k} \rangle , \quad \langle \dots \rangle \equiv  Tr \{\dots e^{-H/T}\}/Tr\{e^{-H/T}\} ,
\label{order}
\end{equation}
\begin{equation}
\bar{\beta}^{'}_{\bf k}= \langle \hat{\beta}^{'}_{\bf k} \rangle_{\mathcal{H}_N}\equiv \frac{\langle \hat{\beta}_{\bf k}\theta(N-\hat{n}) \rangle}{P_N} , \quad P_N = \langle \theta(N-\hat{n}) \rangle ,
\label{order'}
\end{equation}
\begin{equation}
G^{j_2\tau_2{\bf k_2}}_{j_1\tau_1{\bf k_1}}= -\langle T_{\tau} \tilde{\beta}_{j_1\tau_1{\bf k_1}}\tilde{\bar{\beta}}_{j_2\tau_2{\bf k_2}} \rangle ,
\label{Green}
\end{equation}
\begin{equation}
G^{'j_2\tau_2{\bf k_2}}_{j_1\tau_1{\bf k_1}}= -\langle T_{\tau} \tilde{\beta'}_{j_1\tau_1{\bf k_1}}\tilde{\bar{\beta'}}_{j_2\tau_2{\bf k_2}} \rangle /P_N .
\label{Green'}
\end{equation}   
Here $\langle \dots \rangle$ or $\langle \dots \rangle_{\mathcal{H}_N}$ mean the averages over the unconstrained or $\theta(N-\hat{n})$-cutoff excited-states Hilbert spaces $\mathcal{H}$ or ${\mathcal{H}_N}$, respectively. For $\theta(N-\hat{n})$-cutoff operators, like in Eqs. (\ref{order'}) and (\ref{Green'}), they differ only by a factor $P_N$ equal to a cumulative probability of a total occupation of the excited states in the space $\mathcal{H}$ to not exceed the number of particles $N$. $T_{\tau}$ means a standard Matsubara $\tau$-ordering. The indexes $j_{1}=1, 2$ and $j_{2}=1, 2$ enumerate $2\times 2$-matrix of the normal and anomalous Green's functions, since we set $\tilde{A}_{1\tau}\equiv \tilde{A}_{\tau}$ and $\tilde{A}_{2\tau}\equiv \tilde{\bar{A}}_{\tau}$ (a Matsubara-conjugated operator) for any operator $\tilde{A}_{\tau}$.

    Both canonical Green's functions $G$ and $G'$ in Eqs. (\ref{Green})-(\ref{Green'}) are different from the grand-canonical-ensemble ones, employed in a usual Beliaev-Popov theory \cite{AGD,HohenbergMartin,Kondor1974,FetterWalecka,Anderson1984,Shi,Shlyapnikov1998,PitString,RevModPhys2004,ProukakisTutorial2008,LLV,LL} on the unconstrained Fock space $\mathcal{H}^{(0)}$. 
    
    The exact Hamiltonian in Eq. (\ref{Hint}) consists of 7 terms:
\begin{equation}
  H_{N}^{'}= \frac{u_0}{2}N(N-1)\theta(N-\hat{n})+V_{1,1}+V_2+V_2^{\dagger}+V_3+V_3^{\dagger}+V_4, 
\label{Hint7}
\end{equation}
\noindent $V_{1,1}= \frac{N-\hat{n}}{V}\theta(N-\hat{n}) \int \int U({\bf r_1}-{\bf r_2})\hat{\beta}_{\bf r_2}^{\dagger}\hat{\beta}_{\bf r_1} d^3 r_1 d^3 r_2,$

\noindent $V_2 = \frac{\hat{Q}}{2V}\theta(N-\hat{n}) \int \int U({\bf r_1}-{\bf r_2})\hat{\beta}_{\bf r_2}\hat{\beta}_{\bf r_1} d^3 r_1 d^3 r_2,$

\noindent $V_3 = \frac{\sqrt{N-\hat{n}}}{\sqrt{V}}\theta(N-\hat{n}) \int \int U({\bf r_1}-{\bf r_2})\hat{\beta}_{\bf r_2}^{\dagger}\hat{\beta}_{\bf r_2}\hat{\beta}_{\bf r_1} d^3 r_1 d^3 r_2,$

\noindent $V_4 = \frac{1}{2}\theta(N-\hat{n}) \int \int (U-u_0)\hat{\beta}_{\bf r_1}^{\dagger}\hat{\beta}_{\bf r_2}^{\dagger}\hat{\beta}_{\bf r_2}\hat{\beta}_{\bf r_1} d^3 r_1 d^3 r_2;$

\noindent $u_{\bf k}=\int_V U({\bf r})e^{-i{\bf kr}}d^3r/V, \ \hat{Q}=\sqrt{(N-\hat{n})(N-1-\hat{n})}$. 

\noindent In the consistent microscopic theory the Heisenberg equation of motion $\partial \tilde{\beta}_x /\partial \tau = [\tilde{\beta}_x ,H]$ contains a commutator
\begin{equation}
[\tilde{\beta}_x , \tilde{H}_{N\tau}^{'}] = (\tilde{H}_{(N-1)\tau}^{'}-\tilde{H}_{N\tau}^{'})\tilde{\beta}_x +\tilde{A} - \int_V \tilde{A}\frac{d^3 r}{V},
\label{commutator}
\end{equation}
$\tilde{A}= \theta(N-1-\tilde{n}_{\tau})\int U({\bf r'}-{\bf r}) \tilde{\bar{\Psi}}_{(N-1)x'} [\tilde{\Psi}_{(N-1)x'}\tilde{\Psi}_{Nx}$ $+ \tilde{\Psi}_{(N-1)x}\tilde{\Psi}_{Nx'}] \frac{d^3 r'}{2},$
with an operator $\tilde{H}_{(N-1)\tau}^{'}-\tilde{H}_{N\tau}^{'}$, which is incorrectly replaced by a c-number chemical potential $\mu$ in the usual approach. The Hamiltonian $H_{N}^{'}$ turns into the usual Beliaev-Popov one if one replaces the operators $\theta(N-\hat{n})$ and $\sqrt{N-\hat{n}}$ by c-numbers. Such replacement is wrong near the critical $\lambda$-point due to very large critical fluctuations, provided by those operators.

\subsection{3.2. Exact equations for the order parameter and Green's functions of the unconstrained excitations}

    First of all, let us exploit a remarkable fact that, after the first step of the many-body Hilbert space reduction in Eq. (\ref{reduction}) and due to the isomorphism in Eq. (\ref{alpha'=beta'}), we can work in the unconstrained excited-states Hilbert space $\mathcal{H}$, where the standard Matsubara (or Feynman) diagram technique and the exact Dyson equation are valid. Thus, we immediately get the exact microscopic equations for the unconstrained coherent order parameter and Green's functions via a total irreducible self-energy $\Sigma^{j_2x_2}_{j_1x_1}$, as it is implied for the Dyson-type equations. The result is
\begin{equation}
\bar{\beta}_{jx}= \check{G}^{(0)}[\check{\Sigma}[\bar{\beta}_{jx}]],
\label{betaEq}
\end{equation}
\begin{equation}
G_{j_1x_1}^{j_2x_2}+ \bar{\beta}_{j_1x_1}\bar{\beta}^{*}_{j_2x_2} = G_{j_1x_1}^{(0)j_2x_2} + \check{G}^{(0)}[\check{\Sigma}[G_{j_1x_1}^{j_2x_2}]]. 
\label{GEq}
\end{equation}
Here the integral operators $\check{\Sigma}$ or $\check{G}^{(0)}$, applied to any function $f_{jx}$, stand for a convolution of that function $f_{jx}$ over the variables $j, \tau, {\bf r}$ with the total irreducible self-energy $\Sigma$ or a free propagator $G^{(0)}$, respectively:
\begin{equation}
\check{K}[f_{jx}]\equiv \sum_{j'=1}^2 \int K^{j'x'}_{jx} f_{j'x'} d^4 x' \quad \text{for} \ \check{K}=\check{\Sigma}, \check{G}^{(0)},
\label{KEq}
\end{equation}
where $x=\{\tau, {\bf r}\}$ and $\int \dots d^4 x \equiv \int_0^{1/T} \int_V \dots d^3 r d\tau$. The free propagator, that is the zeroth-order Green's function, is determined by a thermal average with the free Hamiltonian in Eq. (\ref{H_0}),
\begin{equation}
G_{j_1x_1}^{(0)j_2x_2}= -\langle T_{\tau} \tilde{\beta}_{j_1x_1} \tilde{{\bar\beta}}_{j_2x_2} \rangle_0, \quad \langle \dots \rangle_0 \equiv  \frac{Tr \{\dots e^{-H_0/T}\}}{Tr\{e^{-H_0/T}\}}.
\label{G0}
\end{equation}

   By means of a known for a box, inverse to $\check{G}^{(0)}$ operator \cite{FetterWalecka}, one can rewrite Eqs. (\ref{betaEq}) and (\ref{GEq}) in a more familiar differential form:
\begin{equation}
\left[ \frac{\hbar \partial}{\partial \tau}+(-1)^j \frac{\hbar^2\nabla_{\bf r}^2}{2M} \right] \bar{\beta}_{jx}= -\sum_{j'=1}^2 \int \Sigma^{j'x'}_{jx} \bar{\beta}_{j'x'} d^4 x' ,
\label{betaEqD}
\end{equation}
$\left[ \frac{\hbar \partial}{\partial \tau_1}+(-1)^{j_1} \frac{\hbar^2\nabla_{\bf r_1}^2}{2M} \right] (G_{j_1x_1}^{j_2x_2} + \bar{\beta}_{j_1x_1}\bar{\beta}^{*}_{j_2x_2})= (-1)^{j_1}\delta_{j_1,j_2}$
\begin{equation}
\times\delta(\tau_1-\tau_2)[\delta({\bf r_1}-{\bf r_2})-\frac{1}{V}] - \sum_{j'=1}^2 \int \Sigma^{j'x'}_{j_1x_1} G_{j'x'}^{j_2x_2} d^4 x' .
\label{GEqD}
\end{equation}

    All effects of the microscopic interaction between particles on the spontaneous symmetry breaking are encoded in the total irreducible self-energy $\Sigma^{j_1x_1}_{j_2x_2}$. Let us stress that already at this step the present microscopic theory of BEC is strongly different from the standard Beliaev-Popov theory. The reason is that the present self-energy essentially differs from the usually used one in the entire critical region since it is calculated for the full, exact interaction Hamiltonian in Eq. (\ref{Hint}) and, hence, includes the additional, constraint interaction and all effects of critical fluctuations. The total irreducible self-energy can be found explicitly from its definition ($\kappa=\{ {\bf k},\tau\}$)
\begin{equation}
\langle T_{\tau} [\tilde{\beta}_{j_1\kappa_1}, \tilde{H}^{'}_{N\tau_1}]\Delta\tilde{\bar{\beta}}_{j_2\kappa_2} \rangle =(-1)^{j_1}\sum_{j=1}^2 \int_0^{\frac{1}{T}} \sum_{{\bf k}\neq 0}\Sigma_{j_1\kappa_1}^{j\kappa}G_{j\kappa}^{j_2\kappa_2} d\tau
\label{self-energy}
\end{equation}
with an exact account for a possible coherent ordering ${\bar{\beta}}_{j\kappa}$. We substitute $\tilde{\beta}_{j\kappa} = \bar{\beta}_{j\kappa}+\Delta\tilde{\beta}_{j\kappa}$ in the left-hand-side operator $\tilde{H}_{N}^{'}$ and use the nonpolynomial diagram technique \cite{PLA2015,KochLasPhys2007,KochJMO2007} together with the exact equations for the partial operator contractions, explained in Sect. 3.3. In another way, we do the first-order, second-order or ladder approximation in interaction, when calculating the left-hand-side average in accord with the main theorem of the diagram technique in an interaction representation for the appropriate free and interaction Hamiltonians: 
\begin{equation}
H_0^{(\Delta)} = \sum_{{\bf k}\neq 0} \varepsilon_{\bf k} \Delta\hat{\beta}_{\bf k}^{\dagger}\Delta\hat{\beta}_{\bf k}, \quad H_N^{'(\Delta)} =H_N^{'} + H_0 - H_0^{(\Delta)}. 
\label{H_0^Delta}
\end{equation}
Of course, a final result does not depend on a way of splitting the total Hamiltonian in a particular pair of the free and interaction Hamiltonians. Both representations $H= H_0 +H'_N$ in Eqs. (\ref{H_0}) and (\ref{Hint}) and $H= H_0^{(\Delta)} +H_N^{'(\Delta)}$ in Eq. (\ref{H_0^Delta}) are equivalent and yield the same total irreducible self-energy (that is $\Sigma= \Sigma^{(\Delta)}$), if one takes into account the corresponding equivalent series of diagrams. A function in the left hand side of Eq. (\ref{GEq}) is known as an exact unconstrained connected Green's function
\begin{equation}
g_{j_1x_1}^{j_2x_2} = G_{j_1x_1}^{j_2x_2}+ \bar{\beta}_{j_1x_1}\bar{\beta}^{*}_{j_2x_2} \equiv -\langle T_{\tau} \Delta \tilde{\beta}_{j_1x_1} \Delta \tilde{\bar{\beta}}_{j_2x_2} \rangle. 
\label{g}
\end{equation}
At the equal times $\tau_1=\tau_2$, it is defined in accord with an anti-normal ordering of the operators $\Delta \tilde{\beta}_{j_1x_1}$ and $\Delta \tilde{\bar{\beta}}_{j_2x_2}$. A reason for the latter is explained in Sect. 3.3.

    In that way, the nonpolynomial diagram technique \cite{PLA2015,KochLasPhys2007,KochJMO2007} yields the left hand side in Eq. (\ref{self-energy}) exactly in a form of its right hand side. The calculations are straightforward, but a full analysis is lengthy and will be presented elsewhere. At its last stage, it is based on a distribution of the noncondensate occupation \cite{PRA2010,PRA2014}, 
\begin{equation}
\rho_n =\int^{\pi}_{-\pi}e^{\phi-iun}\frac{du}{2\pi}, \ \phi= \sum^{\infty}_{m=1}\tilde{\kappa}_m^{(\infty)} \frac{(e^{iu}-1)^m}{m!}. 
\label{rho_n}
\end{equation}
The latter is given by the generating cumulants $\tilde{\kappa}_m^{(\infty)}$. In a presence of an arbitrary coherent ordering $\bar{\beta}_{\bf k}\neq 0$ for a particular case of an ideal gas we find them exactly:
\begin{equation}
\tilde{\kappa}_m^{(\infty)} =(m-1)! \sum_{{\bf k}\neq 0} [\bar{n}_{\bf k}^{m}+m\bar{n}_{\bf k}^{m-1}|\bar{\beta}_{\bf k}|^2], \bar{n}_{\bf k}= (e^{\frac{\varepsilon_{\bf k}}{T}}-1)^{-1}.
\label{rho}
\end{equation}

\subsection{3.3. Fundamental equations for the constrained, physical excitations: The order parameter, Green's functions and partial operator contractions}

    The main problem with the microscopic theory for the true, constrained order parameter and Green's functions, Eqs. (\ref{order'}) and (\ref{Green'}), is that there are no closed equations available for them in the critical region, at least, of the Dyson type. A reason is that they are the averages which include, along with the usual products of the creation and annihilation operators, the nonpolynomial functions, like the ones ($\theta(N-\hat{n})$ or $\sqrt{N-\hat{n}}$), entering the true excitation operators in Eq. (\ref{hatbeta'}) and the exact constraint interaction Hamiltonian in Eq. (\ref{Hint}). The standard diagram methods are not suited for the nonpolynomial averages. 
    
    We find a universal solution to that problem by means of the nonpolynomial diagram technique \cite{PLA2015,KochLasPhys2007,KochJMO2007}. Namely, the true order parameter and Green's functions in Eqs. (\ref{order'}) and (\ref{Green'}) are given by the following explicit formulas
\begin{equation}
\bar{\beta}^{'}_{J}= \langle \tilde{s}_J[\theta (N-\tilde{n}_{\tau_i})] \rangle /P_N , \quad P_N = \langle \theta(N-\hat{n}) \rangle ,
\label{beta'}
\end{equation}
\begin{equation}
G^{'J_2}_{J_1}= -\langle \tilde{s}_{\bar{J}_2}[ \tilde{s}_{J_1}[\theta ]] + \tilde{b}^{J_2}_{J_1}[\theta ] \rangle /P_N .
\label{G'}
\end{equation}
Here the basis partial one- and two-operator contractions, $\tilde{s}_J[f]$ and $\tilde{b}^{J_2}_{J_1}[f]$, are the operator-valued functionals, evaluated for the operator functions $f=\theta (N-\tilde{n}_{\tau_i})$ or $f=\theta \equiv \theta (N-\tilde{n}_{\tau_1})\theta (N-\tilde{n}_{\tau_2})$. We use the short notations for the combined indexes $J= \{ji{\bf k_i}\}$ and $J_l= \{j_l i_l{\bf k_{i_l}}\}$ as well as for an index $\bar{J}_l= \{\bar{j}_li_l{\bf k_{i_l}}\}$ with a complimentary index $\bar{j}_l=3-j_l$; $j_l=1,2$. An index $i=1,2$ (or $i_l$) enumerates different times $\tau_i$ (or $\tau_{i_l}$) in the external operator $\tilde{\beta}_{j\tau_i{\bf k_i}}$ (or $\tilde{\beta}_{j_l\tau_{i_l}{\bf k_{i_l}}}$). 

    The basis partial one-operator contraction $\tilde{s}_J[f]$ of the Matsubara operator $\tilde{\beta}_J$ with an arbitrary function $f(\tilde{n}_{\tau_1},\tilde{n}_{\tau_2})$ of the Matsubara operators $\tilde{n}_{\tau_1}$ and $\tilde{n}_{\tau_2}$ for the total occupation of the excited states, 
\begin{equation}
\tilde{s}_J[f(\tilde{n}_{\tau_1},\tilde{n}_{\tau_2})]\equiv \mathcal{A}_{\tau_i} T_{\tau}\{\tilde{\beta}_J^{c}f^c(\tilde{n}_{\tau_1},\tilde{n}_{\tau_2}) \},
\label{s}
\end{equation}
is defined by an explicit formula via the $f$'s Taylor series:
$$\tilde{s}_J[f(\tilde{n}_{\tau_1},\tilde{n}_{\tau_2})]= T_{\tau} \sum_{m_1=0}^{\infty}\sum_{m_2=0}^{\infty} \sum_{m'_1=0}^{m_1}\sum_{m'_2=0}^{m_2} \frac{f^{(m_1,m_2)}}{m_1!m_2!}$$
\begin{equation}
\times \langle \mathcal{A}_{\tau_i} T_{\tau}\{\tilde{\beta}_J \tilde{n}_{\tau_1}^{m_1-m'_1}\tilde{n}_{\tau_2}^{m_2-m'_2}\} \rangle \tilde{n}_{\tau_1}^{m'_1}\tilde{n}_{\tau_2}^{m'_2},
\label{sTaylor}
\end{equation}
where $f^{(m_1,m_2)}=\partial^{m_1+m_2}f/(\partial^{m_1}\tilde{n}_{\tau_1} \partial^{m_2}\tilde{n}_{\tau_2})|_{\tilde{n}_{\tau_1}=0,\tilde{n}_{\tau_2}=0}$. An examination of the case of two variables $(\tilde{n}_{\tau_1},\tilde{n}_{\tau_2})$ is enough for a generic definition. A symbol $\mathcal{A}_{\tau_i}$ denotes an anti-normal ordering, which prescribes a position of the external operator $\tilde{\beta}_J$ relative to the function $f(\tilde{n}_{\tau_1},\tilde{n}_{\tau_2})$ and does not affect any other operators' positions, which are determined by the $T_{\tau}$-ordering. As is explained in the next Sect. 3.4, this definition implies a sum of all possible partial connected contractions, which were used in the nonpolynomial diagram technique in \cite{PLA2015,KochLasPhys2007,KochJMO2007} and denoted by the superscripts "c" in Eqs. (\ref{s}), (\ref{b}), (\ref{sint}), and (\ref{bint}). Note that, contrary to the usual notations \cite{AGD}, here "c" means only the partial (not full !) contraction.

    The basis partial two-operator contraction $\tilde{b}_{J_1}^{J_2}[f]$ of the two Matsubara operators $\Delta \tilde{\beta}_{J_1}$ and $\Delta \tilde{\bar{\beta}}_{J_2}$ with an arbitrary function $f(\tilde{n}_{\tau_1},\tilde{n}_{\tau_2})$ of the two occupation Matsubara operators $\tilde{n}_{\tau_1}$ and $\tilde{n}_{\tau_2}$,
\begin{equation}
\tilde{b}_{J_1}^{J_2}[f(\tilde{n}_{\tau_1},\tilde{n}_{\tau_2})] \equiv \mathcal{A}_{\tau_{i_1}\tau_{i_2}} T_{\tau} \{ \Delta \tilde{\beta}_{J_1}^{c}\Delta \tilde{\bar{\beta}}_{J_2}^{c} f^c(\tilde{n}_{\tau_1},\tilde{n}_{\tau_2}) \} ,
\label{b}
\end{equation}
is defined by an analytical formula similar to Eq. (\ref{sTaylor}):
$$\tilde{b}_{J_1}^{J_2}[f(\tilde{n}_{\tau_1},\tilde{n}_{\tau_2})]= T_{\tau} \sum_{m_1=0}^{\infty}\sum_{m_2=0}^{\infty} \sum_{m'_1=0}^{m_1}\sum_{m'_2=0}^{m_2} \frac{f^{(m_1,m_2)}}{m_1!m_2!}$$
\begin{equation}
\times \langle \mathcal{A}_{\tau_{i_1}\tau_{i_2}} T_{\tau}\{\Delta \tilde{\beta}_{J_1}\Delta \tilde{\bar{\beta}}_{J_2} \tilde{n}_{\tau_1}^{m_1-m'_1}\tilde{n}_{\tau_2}^{m_2-m'_2}\} \rangle \tilde{n}_{\tau_1}^{m'_1}\tilde{n}_{\tau_2}^{m'_2}.
\label{bTaylor}
\end{equation}
Again, the anti-normal ordering $\mathcal{A}_{\tau_{i_1}\tau_{i_2}}$ prescribes only the positions of the external operators $\Delta \tilde{\beta}_{J_1}$ and $\Delta \tilde{\bar{\beta}}_{J_2}$ relative to the function $f(\tilde{n}_{\tau_1},\tilde{n}_{\tau_2})$ and does not affect any other operators' positions, which are determined by the $T_{\tau}$-ordering. The partial operator contractions with the other than anti-normal orderings also exist and could be calculated, but the anti-normally ordered ones in Eqs. (\ref{s}) and (\ref{b}) are more convenient since the true, constrained creation and annihilation operators in Eqs. (\ref{alpha'}) and (\ref{hatbeta'}) are anti-normally ordered relative to their step-functions $\theta$. 

    The exact closed difference (recurrence) equations for the basis partial operator contractions $\tilde{s}_{J}[f(\{ m_{J'} \})]$ and $\tilde{b}^{J_2}_{J_1}[f(\{ m_{J'} \})]$ for an arbitrary function $f(\{ m_{J'} \}) =f(\tilde{n}_{\tau_1}+N+1-m_1, \tilde{n}_{\tau_2}+N+1-m_2)$, where a set $\{ m_{J'}\}\equiv \{ m_{i'}\}$ consists of the integers $m_1$ and $m_2$, are derived in the subsequent Sect. 3.4. Those equations have the following universal form:
\begin{equation}
\tilde{s}_{J}[f(\{ m_{J'} \})]= \bar{\beta}_J f(\{ m_{J'} \})+ g_J^{J'} \Delta_{m_{J'}} \tilde{s}_{J'}[f(\{ m_{J'} \})],
\label{1-contraction}
\end{equation}
$$\tilde{b}^{J_2}_{J_1}[f(\{ m_{J'} \})] = -g_{J_1}^{J_2}f(\{ m_{J'} \}) - g_{J_1}^{J'} g_{J'}^{J_2} \Delta_{m_{J'}}f(\{ m_{J'} \})$$
\begin{equation}
+g_{J_1}^{J'_1} g_{J'_2}^{J_2} \Delta_{m_{J'_1}} \Delta_{m_{J'_2}} \tilde{b}^{J'_2}_{J'_1}[f(\{ m_{J'} \})].
\label{2-contraction}
\end{equation}

   A matrix $g_J^{J'}$, which enters these equations as the coefficients, is given by the solution of Eq. (\ref{GEq}) for the exact unconstrained connected Green's function in Eq. (\ref{g}). For a special case of the equal times $\tau_i=\tau_{i'}$ it is defined in accord with an anti-normal ordering of the operators $\Delta \tilde{\beta}_J$ and $\Delta \tilde{\bar{\beta}}_{J'}$. The latter is dictated by the anti-normal ordering in the definition of the partial operator contractions in Eqs. (\ref{s})-(\ref{bTaylor}). In Eqs. (\ref{1-contraction})-(\ref{2-contraction}), a symbol $\Delta_{m_{J'}}$ means a partial difference operator \cite{DE-Elaydi} ($\Delta_{m_1} f({m_1,m_2})=f({m_1+1,m_2})-f({m_1,m_2})$ and $\Delta_{m_2} f({m_1,m_2})=f({m_1,m_2+1})-f({m_1,m_2})$) and we assume an Einstein's summation over the repeated indexes $J', J'_1, J'_2$. The sums run over $j'=1,2$ and all different arguments $\tilde{n}_{{\bf k'_{i'}}\tau'_{i'}}$ of the function $f$, enumerated by $i'=1,2$, ${\bf k'_{i'}}\neq 0$, for $J'$ and similarly for $J'_1, J'_2$. 

    The equations (\ref{1-contraction}) and (\ref{2-contraction}) are the linear systems of the integral equations over the wave vector variables and the discrete (recurrence) equations over variables $m_1$ and $m_2$ with the well-known methods of solution \cite{PDE-Cheng,DE-Agarwal,DE-Elaydi}, e.g., via a Z-transform (a discrete analog of a Laplace transform), a characteristic function (a solution for $f=\exp(iu_1\tilde{n}_{\tau_1}+iu_2\tilde{n}_{\tau_2})$ with a subsequent Fourier transform), or a direct recursion. The operator contractions in Eqs. (\ref{beta'})-(\ref{G'}) are given by those solutions at $m_1=m_2=N+1$. 

    Finally, their average in Eqs. (\ref{beta'})-(\ref{G'}) amounts to the averages like $\langle f(\tilde{n}_{\tau_1},\tilde{n}_{\tau_2}) \rangle$, that is reduced to a calculation of a joint probability distribution $\rho_{n_1,n_2}$ of the noncommuting operators $\tilde{n}_{\tau_1}$ and $\tilde{n}_{\tau_2}$. The latter can be done similar to \cite{PRA2010,PRA2014} via its characteristic function $\langle T_{\tau} \exp(iu_1\tilde{n}_{\tau_1}+iu_2\tilde{n}_{\tau_2}) \rangle$, which is equal to
\begin{equation}
\Theta= \prod_{{\bf k}\neq 0} \frac{e^{\frac{|\bar{\beta}_{\bf k}|^2 [(z_{\bf k}-1)(z_1z_2-1)-(z_{\bf k}-e^{\tau \varepsilon_{\bf k}})(1-e^{-\tau \varepsilon_{\bf k}})(z_1-1)(z_2-1)]}{z_{\bf k}-z_1z_2}}}{[(z_{\bf k}-z_1z_2)/(z_{\bf k}-1)]}
\label{2Dchar-function}
\end{equation}
for any $\bar{\beta}_{\bf k}$ in the zeroth order in interaction; $\tau=\tau_1-\tau_2, z_{\bf k}=e^{\frac{\varepsilon_{\bf k}}{T}}, z_j =e^{iu_j}, j=1,2$. The result in Eq. (\ref{2Dchar-function}), together with its particular case for $u_2=0$ in Eqs. (\ref{rho_n})-(\ref{rho}), provides a basis for a perturbation analysis. The macroscopic wave function (the coherent order parameter) of the excitations' condensate $\bar{\beta}_{\bf r}\neq 0$, entering Eqs. (\ref{GEq}), (\ref{GEqD}), and (\ref{beta'})-(\ref{2-contraction}), makes the quasiparticles stable and can be found from Eq. (\ref{betaEq}) or Eq. (\ref{betaEqD}).

\subsection{3.4. The nonpolynomial diagram technique: Derivation of the exact recurrence equations for the basis partial operator contractions}

    We start a derivation of the equations for the operator contractions with a formulation of the nonpolynomial diagram technique for them in the interaction representation with the free Hamiltonian $H_0$ in Eq. (\ref{H_0}) or $H_0^{(\Delta)}$ in Eq. (\ref{H_0^Delta}). Let $\hat{A}_{\tau}$ denotes an interaction representation of any operator $\hat{A}$, that is $\hat{A}_{\tau}=\exp(\tau H_0)\hat{A}\exp(-\tau H_0)$ or $\hat{A}_{\tau}=\exp(\tau H_0^{(\Delta)})\hat{A}\exp(-\tau H_0^{(\Delta)})$. 
    
    First, let us consider an equivalent definition of the basis partial one-operator contraction in the interaction representation. According to the main theorem of the diagram technique \cite{AGD,FetterWalecka}, a thermal average of the Matsubara operator $T_{\tau}\{\tilde{\beta}_{J_1} f(\tilde{n}_{\tau_{i_1}},\tilde{n}_{\tau_{i_2}}) \}$ over a statistical distribution $\exp(-\frac{H}{T})$ with a full Hamiltonian $H$ is equal to a thermal average of the interaction-representation operator $T_{\tau}\{\hat{\beta}_{J_1} f(\hat{n}_{\tau_{i_1}},\hat{n}_{\tau_{i_2}}) S(\frac{1}{T}) \}$ over a statistical distribution $\exp(-\frac{H_0}{T})$ with a free Hamiltonian $H_0$. Here a symbol $S(\frac{1}{T})$ stands for a standard Matsubara S-matrix
\begin{equation}
S\Big(\frac{1}{T}\Big) = T_{\tau} \exp \Big( -\int_0^{1/T}\hat{H}^{'}_{N\tau} d\tau \Big).
\label{Smatrix}
\end{equation}
We omit a normalization factor $\langle S(\frac{1}{T})\exp (-H_0/T) \rangle_0$, which is canceled since, as usual, we consider only the connected diagrams. Hence, in the interaction representation the basis partial one-operator contraction 
\begin{equation}
\hat{s}_{J_1}[f(\hat{n}_{\tau_{i_1}},\hat{n}_{\tau_{i_2}})]\equiv \mathcal{A}_{\tau_{i_1}} T_{\tau}\{\hat{\beta}_{J_1}^{c}f^c(\hat{n}_{\tau_{i_1}},\hat{n}_{\tau_{i_2}}) S\Big(\frac{1}{T}\Big) \},
\label{sint}
\end{equation}
corresponding to the Matsubara one in Eqs. (\ref{s})-(\ref{sTaylor}), is equal to $\bar{\beta}_{J_1}f(\hat{n}_{\tau_{i_1}},\hat{n}_{\tau_{i_2}})$ plus a sum of all operators obtained from the operator $\mathcal{A}_{\tau_{i_1}} T_{\tau} \{ \Delta \hat{\beta}_{J_1} f(\hat{n}_{\tau_{i_1}},\hat{n}_{\tau_{i_2}}) S(\frac{1}{T}) \}$
by all possible partial connected contractions via the 1D-connected diagrams, which start from the external operator $\Delta \hat{\beta}_{J_1}$, connect a series of vertices 
\begin{equation}
\hat{n}_{J'_l}= (\Delta \hat{\bar{\beta}}_{J'_l}+\bar{\beta}^{*}_{J'_l}) (\Delta \hat{\beta}_{J'_l}+\bar{\beta}_{J'_l}) 
\label{nJ}
\end{equation}
inside the function $f(\hat{n}_{\tau_{i_1}},\hat{n}_{\tau_{i_2}})$ via the elementary contractions in Eq. (\ref{DeltaG0}), and end at one of the annihilation or creation operators $\Delta \hat{\bar{\beta}}_{J'}$ inside the function $f(\hat{n}_{\tau_{i_1}}, \hat{n}_{\tau_{i_2}})$. An elementary contraction is a negative connected free Green's function
\begin{equation}
\langle T_{\tau} \Delta \hat{\beta}_{J_1} \Delta \hat{\bar{\beta}}_{J_2} \rangle_0 \equiv -g_{J_1}^{(0)J_2}= - G_{J_1}^{(0)J_2} - \bar{\beta}_{J_1} \bar{\beta}^{*}_{J_2}.
\label{DeltaG0}
\end{equation}

    An equivalent definition of the basis partial two-operator contraction in the interaction representation,
\begin{equation}
\hat{b}_{J_1}^{J_2}[f]\equiv \mathcal{A}_{\tau_{i_1}\tau_{i_2}} T_{\tau}\{\Delta \hat{\beta}_{J_1}^{c}\Delta \hat{\bar{\beta}}_{J_2}^{c} f^c(\hat{n}_{\tau_{i_1}},\hat{n}_{\tau_{i_2}}) S\Big(\frac{1}{T}\Big) \},
\label{bint}
\end{equation}
which corresponds to the Matsubara one defined in Eqs. (\ref{b})-(\ref{bTaylor}), is similar. Namely, the $\hat{b}_{J_1}^{J_2}[f]$ is equal to a sum of all operators obtained from the operator $\mathcal{A}_{\tau_{i_1}\tau_{i_2}} T_{\tau} \{ \Delta \hat{\beta}_{J_1} \Delta \hat{\bar{\beta}}_{J_2} f(\hat{n}_{\tau_{i_1}},\hat{n}_{\tau_{i_2}}) S(\frac{1}{T}) \}$
by all possible partial connected contractions via the 1D-connected diagrams, which start from one external operator $\Delta \hat{\beta}_{J_1}$, connect a series of vertices $\hat{n}_{J'_l}= (\Delta \hat{\bar{\beta}}_{J'_l}+\bar{\beta}^{*}_{J'_l}) (\Delta \hat{\beta}_{J'_l}+\bar{\beta}_{J'_l})$ inside the function $f(\hat{n}_{\tau_{i_1}},\hat{n}_{\tau_{i_2}})$ via the elementary contractions in Eq. (\ref{DeltaG0}), and end at another external operator $\Delta \hat{\bar{\beta}}_{J_2}$.

    An elementary, generic block of such contractions is a partial contraction of the operator $\Delta \hat{\beta}_{{\bf k}\tau}$ with the $m$-th power of the total excited-states occupation operator $\hat{n}^m_{\tau'}$. Let us make just one elementary contraction (\ref{DeltaG0}) of the external operator $\Delta \hat{\beta}_J$ with each creation ($\Delta \hat{\bar{\beta}}_{{\bf k'}\tau'}$) or annihilation ($\Delta \hat{\beta}_{{\bf k'}\tau'}$) operators from each factor $\hat{n}_{\tau'}$ (see Eqs. (\ref{N}) and (\ref{nJ})), located at the positions $m'=1,\dots,m$ in a sequence of $m$ factors in the $\hat{n}_{\tau'}^m$. The result of that one-step contraction operation is a following sum:
\begin{equation}
T_{\tau} \{ \Delta \hat{\beta}_{{\bf k}\tau} \hat{n}^m_{\tau'} \} \to \sum_{{\bf k'}\neq 0} \sum_{m'=1}^m \Big[ \langle T_{\tau} \Delta \hat{\beta}_{{\bf k}\tau} \Delta \hat{\bar{\beta}}_{{\bf k'}\tau'} \rangle_0 C_1(m') 
\label{element}
\end{equation}
$$+ \langle T_{\tau} \Delta \hat{\beta}_{{\bf k}\tau} \Delta \hat{\beta}_{{\bf k'}\tau'} \rangle_0 C_2(m') \Big],$$
where the operators $C_1(m')=\hat{n}_{\tau'} \dots \hat{n}_{\tau'} \hat{\beta}_{{\bf k'}\tau'} \hat{n}_{\tau'} \dots \hat{n}_{\tau'}$ and $C_2(m')=\hat{n}_{\tau'} \dots \hat{n}_{\tau'} \hat{\bar{\beta}}_{{\bf k'}\tau'} \hat{n}_{\tau'} \dots \hat{n}_{\tau'}$ depend on a position number $m'$, at which the remaining after that one-step contraction operators $\hat{\beta}_{{\bf k'}\tau'}$ and $\hat{\bar{\beta}}_{{\bf k'}\tau'}$ are located in the sequence of the $\hat{n}_{\tau'}$-factors. The canonical commutation relations yield $C_1(m')= \hat{\beta}_{{\bf k'}\tau'}(\hat{n}_{\tau'}-1)^{m'-1} \hat{n}^{m-m'}_{\tau'}$ and $C_2(m')= \hat{n}^{m'-1}_{\tau'} (\hat{n}_{\tau'}-1)^{m-m'} \hat{\bar{\beta}}_{{\bf k'}\tau'}$. The corresponding sums over $m'$ are reduced to a discrete difference $\Delta_n \hat{n}^m_{\tau'} \equiv \hat{n}^m_{\tau'} -(\hat{n}_{\tau'}-1)^m$, namely, $\sum_{m'=1}^m C_1(m')= \hat{\beta}_{{\bf k'}\tau'} \Delta_n \hat{n}^m_{\tau'}$ and $\sum_{m'=1}^m C_2(m')= [\Delta_n \hat{n}^m_{\tau'}] \hat{\bar{\beta}}_{{\bf k'}\tau'}$. 

    Then, let us dress the free propagators in Eq. (\ref{element}) into the exact unconstrained connected Green's functions (\ref{g}) by including all appropriate diagrams, coming from the S-matrix in Eq. (\ref{Smatrix}). In this way, we conclude that the considered one-step elementary contraction is a transformation of the original operator $T_{\tau} \{ \Delta \hat{\beta}_J \hat{n}^m_{\tau'} \}$ into a convolution of its discrete-difference counterpart $T_{\tau} \{ \hat{\beta}_{J'} \Delta_n \hat{n}^m_{\tau'} \}$ with the exact unconstrained connected Green's function $g_J^{J'}$, that is
\begin{equation}
T_{\tau} \{ \Delta \hat{\beta}_J \hat{n}^m_{\tau'} \} \to -\sum_{{\bf k'}\neq 0} \sum_{j'=1}^2 g_J^{j'\tau'{\bf k'}} T_{\tau} \{ \hat{\beta}_{j'{\bf k'}\tau'} \Delta_n \hat{n}^m_{\tau'} \}.
\label{recursion}
\end{equation}

    It follows from the definition in Eq. (\ref{sint}), that if one supplements this first step by all other required contractions, then the operator $T_{\tau} \{ \hat{\beta}_{J'} \Delta_n \hat{n}^m_{\tau'} \}$ in the right hand side of Eq. (\ref{recursion}) will become a basis partial operator contraction $\hat{s}_{J'}[\Delta_n \hat{n}^m_{\tau'}]$. Therefore, taking into account an obvious term $\bar{\beta}_J \hat{n}^m_{\tau'}$ and combining all terms of the Taylor series, we immediately arrive to the final Eq. (\ref{1-contraction}) for the basis partial one-operator contraction. 
    
    A derivation of the final Eq. (\ref{2-contraction}) for the basis partial two-operator contraction also immediately follows from a property of the one-step contraction in Eq. (\ref{recursion}) to reproduce an original operator as its difference. Namely, the second term in Eq. (\ref{2-contraction}) comes from an additional contraction of the operator $\hat{\beta}_{j'{\bf k'}\tau'}$ from the right hand side of Eq. (\ref{recursion}) with the second external operator $\Delta \hat{\bar{\beta}}_{J_2}$ of $\hat{b}_{J_1}^{J_2}[f]$ in its definition (\ref{bint}). The third term in Eq. (\ref{2-contraction}) is a result of a parallel application of the one-step contraction in Eq. (\ref{recursion}) to both external operators $\Delta \hat{\beta}_{J_1}$ and $\Delta \hat{\bar{\beta}}_{J_2}$ in $\hat{b}_{J_1}^{J_2}[f]$. The first term in Eq. (\ref{2-contraction}) is a result of a direct contraction between the external operators.

\subsection{3.5. Discussion of the exact microscopic equations for the Bose-Einstein condensation} 

\noindent The derived fundamental equations (\ref{betaEq})-(\ref{GEq}), (\ref{beta'})-(\ref{G'}) and (\ref{1-contraction})-(\ref{2-contraction}) consistently account for the critical fluctuations as well as for the effects of the inter-particle interaction on the condensate and quasiparticles. These equations have a perfectly canonical and proper matrix structure. The unconstrained order parameter $\bar{\beta}_J$ and Green's function $G^{J_2}_{J_1}$ form the wavevector-dependent 4-component vector and $4\times 4$-matrix, respectively, since the combined index $J= \{ji{\bf k_i} \}$ includes the Nambu index $j=1,2$ and the time $\tau_i$ index $i=1,2$, each with the two values. The $\bar{\beta}_J$ and $G^{J_2}_{J_1}$ are the solutions of the canonical Dyson-type equations (\ref{betaEq})-(\ref{GEq}) or (\ref{betaEqD})-(\ref{GEqD}), which are the integral equations over the wave vector variable, and are determined by the $4\times 4$-matrices of the free propagator $G^{(0)J_2}_{J_1}$ in Eq. (\ref{G0}) and the total irreducible self-energy $\Sigma^{J_2}_{J_1}$, defined by a two-body interaction in Eq. (\ref{self-energy}).

    In their turn, these solutions $\bar{\beta}_J$ and $G^{J_2}_{J_1}$ enter the inhomogeneous linear integral matrix equations (\ref{1-contraction})-(\ref{2-contraction}) for the basis partial one- and two-operator contractions $\tilde{s}_{J}[f(\{ m_1,m_2 \})]$ (4-vector) and $\tilde{b}^{J_2}_{J_1}[f(\{ m_1,m_2 \})]$ ($4\times 4$-matrix) via the source terms and the coefficients. The coefficients are given by the unconstrained connected Green's function $4\times 4$-matrix $g^{J_2}_{J_1}=G^{J_2}_{J_1}+ \bar{\beta}_{J_1}\bar{\beta}^*_{J_2}$ in Eq. (\ref{g}) (cf. a usual Nambu's $2\times 2$-matrix of the normal and anomalous Green's functions). In addition to being matrix and integral, the latter equations (\ref{1-contraction})-(\ref{2-contraction}) are the partial difference equations (a discrete analog of the partial differential equations) over the variables $m_1$ and $m_2$. The difference equations are required by a discrete, quantum nature of the excitations.

   The revealed nontrivial universal structure of these fundamental equations is uniquely prescribed by the microscopic physics of the critical phenomena because these equations are not the model, approximate, or phenomenological ones, but the exact equations. It is immediate to generalize them to any trapping potentials and boundary conditions. They open a way to solve the long-standing problem of the BEC and other phase transitions \cite{AGD,HohenbergMartin,Kondor1974,FetterWalecka,Anderson1984,Shi,Shlyapnikov1998,PitString,RevModPhys2004,ProukakisTutorial2008,LLV,LL,PatPokr,Fisher1986}, including a restricted canonical ensemble problem \cite{HohenbergMartin}, and describe numerous modern laboratory and numerical experiments on the critical phenomena in BEC of the mesoscopic systems \cite{BrownTwissOnBECThresholdNaturePhys2012,Hadzibabic2013NaturePhys,Hadzibabic2013,EsslingerCriticalBECScience2007,TwinAtomBeamNaturePhys2011,Dalibard2012NaturePhys,Dalibard2010,RaizenTrapControl,Blume,BECinterferometerOnChip2013,EntanglementOnChipNature2010,chipBECNature2001,HeDroplets,Reppy}. 

   A principle difference of these microscopic equations from the usual Gross-Pitaevskii and Beliaev-Popov equations comes from a presence of the operator functions $\theta(N-\hat{n})$ and $\sqrt{N-\hat{n}}$ in the true Hamiltonian (\ref{Hint}). Also, it comes from the symmetry-constrained nature of the actual excitation operators $\hat{\beta}^{'}_{\bf k}=\hat{\beta}_{\bf k}\theta(N-\hat{n})$, which determine the true order parameter and Green's functions in Eqs. (\ref{order'}) and (\ref{Green'}). A replacement of those operator functions by the c-numbers $\theta(N-\hat{n})\approx 1, \sqrt{N-\hat{n}}\approx \bar{n}_0$, that is a mean-field approximation, becomes valid only far outside the critical region, at low enough temperatures, when the ordered phase is fully formed and the order parameter is much larger than its fluctuations. So, the exact microscopic equations asymptotically turn into and, hence, justify the usual Gross-Pitaevskii and Beliaev-Popov ones only far outside the critical region.
   
        Note that, within the particular case of the homogeneous BEC in a box with the periodic boundary conditions, the usually used Beliaev-Popov equations describe BEC as a smooth change of the stable quasiparticles, dressed by a Bogoliubov coupling via a condensate occupation and possessing the inter-mode correlations (the anomalous Green's functions) without any own coherence. In that picture, a coherent superfluid flow is generated by some external sources or boundary motion, but an appearance of the coherent macroscopic order parameter $\bar{\beta}^{'}_{\bf r}$ is excluded by the Hugenholtz-Pines ($T=0$) or Hohenberg-Martin ($T\neq 0$) condition \cite{HohenbergMartin} $(\Sigma^{1{\bf k}}_{1{\bf k}} -\Sigma^{1{\bf k}}_{2{\bf -k}})|_{{\bf k}=0} =\mu$, that ensures a stable gapless spectrum of quasiparticles. That usual mechanism of BEC is similar to the one in an ideal gas, being just its dressed-by-interaction version. It differs from the outlined general mechanism, allowing an instability of restructuring of the condensate by $\bar{\beta}^{'}_{\bf r}\neq 0$, that is a symmetry breaking (cf. a density wave in a superfluid \cite{Anderson1984}). A detailed analysis of Eqs. (\ref{betaEq})-(\ref{GEq}), (\ref{beta'})-(\ref{G'}) and (\ref{1-contraction})-(\ref{2-contraction}) will be given elsewhere.

   The major point is that the microscopic theory is valid both inside and outside the critical region. In particular, it is capable of a microscopic calculation of the critical exponents and functions as well as other parameters of the BEC phase transition near the $\lambda$-point for an actual, canonical-ensemble mesoscopic system, as opposed to a grand-canonical-ensemble bulk-limit model, usually studied in the phenomenological renormalization-group approach \cite{RevModPhys2004,LLV,LL,PatPokr,Fisher1986,Goldenfeld,Vicari2002}.

    There is another known limit of this microscopic theory. Namely, for a vanishing interaction, when $H_{N}^{'} \equiv 0$ and $\bar{\beta}_{\bf k} \equiv 0$ in Eqs. (\ref{rho}) and (\ref{2Dchar-function}), it coincides with a solution for the critical fluctuations in the BEC of a mesoscopic ideal gas \cite{PRA2010,PRA2014}, which is its zeroth-order approximation.
    
     In short, all these advances come from the correct account, first, of the Noether's symmetry constraints of the many-body Hilbert space and, second, of the related properties of the true excitations in a mesoscopic system.

\section{4. The rigorous microscopic theory of the magnetic phase transitions in a lattice of spins}

   The microscopic theory of phase transitions, described above, is truly universal. Let us demonstrate that fact by formulating the microscopic theory of the magnetic phase transitions in a form, which is completely similar to the BEC one, developed in Sect. 3. The only important difference is that in the case of the magnetic phase transition there are many local constraints (\ref{spin}) for each of N spins in a lattice, instead of just one global particle-number constraint (\ref{N}) in the case of BEC in a gas.

\subsection{4.1. Spin bosons versus true spin excitations in a constrained Hilbert space and the Heisenberg Hamiltonian in a Holstein-Primakoff representation}

   Let us consider a cubic lattice of $N$ interacting immovable spins in a box with a volume $V=L^3$ and the periodic boundary conditions. A magnitude of each spin is the same $s \geq 1/2$. The lattice sites of the spins are enumerated by a position vector ${\bf r}$. According to a well-known Holstein-Primakoff representation \cite{HolsteinPrimakoff}, worked out also by Schwinger \cite{Schwinger1965} (see \cite{Sakurai}), each spin is a system of the two spin bosons, which are the harmonic oscillators constrained to have a fixed total occupation
\begin{equation}
\hat{n}_{0{\bf r}}+\hat{n}_{\bf r} =2s; \quad \hat{n}_{\bf r}=\hat{a}^{\dagger}_{\bf r} \hat{a}_{\bf r}, \quad \hat{n}_{0{\bf r}}=\hat{a}^{\dagger}_{0{\bf r}} \hat{a}_{0{\bf r}}.
\label{spin}
\end{equation}
The creation ($\hat{a}^{\dagger}_{\bf r}$, $\hat{a}^{\dagger}_{0{\bf r}}$) and annihilation ($\hat{a}_{\bf r}$, $\hat{a}_{0{\bf r}}$) operators obey the Bose canonical commutation relations: $[\hat{a}_{\bf r}, \hat{a}_{\bf r'}^{\dagger }]=\delta_{{\bf r},{\bf r'}}$, $[\hat{a}_{0{\bf r}}, \hat{a}_{0{\bf r'}}^{\dagger }]=\delta_{{\bf r},{\bf r'}}$, and all $({\bf r})$-operators commute with all $(0{\bf r'})$-operators. A vector spin operator $\hat{\bf S}_{\bf r}$ at a site ${\bf r}$ is given by its components as follows
\begin{equation}
\hat{S}^x_{\bf r}=\frac{\hat{a}^{\dagger}_{0{\bf r}}\hat{a}_{\bf r} + \hat{a}^{\dagger}_{\bf r}\hat{a}_{0{\bf r}}}{2}, \ \hat{S}^y_{\bf r}=\frac{\hat{a}^{\dagger}_{0{\bf r}}\hat{a}_{\bf r}- \hat{a}^{\dagger}_{\bf r}\hat{a}_{0{\bf r}}}{2i}, \ \hat{S}^z_{\bf r}= s - \hat{a}^{\dagger}_{\bf r}\hat{a}_{\bf r}.
\label{spinvector}
\end{equation}
The spin raising and lowering operators are equal to
\begin{equation}
\hat{S}^{+}_{\bf r}\equiv \hat{S}^x_{\bf r}+i\hat{S}^y_{\bf r} =\hat{a}^{\dagger}_{0{\bf r}}\hat{a}_{\bf r}, \quad \hat{S}^{-}_{\bf r}\equiv \hat{S}^x_{\bf r}-i\hat{S}^y_{\bf r} = \hat{a}^{\dagger}_{\bf r}\hat{a}_{0{\bf r}}.
\label{spin+-}
\end{equation}

    Similar to Sect. 3.1, Eq. (\ref{reduction}), we make the two-step reduction of an unconstrained Fock space $\mathcal{H}^{(0)}_{0{\bf r},{\bf r}}$, which spans all Fock states $|n_{\bf r}\rangle \otimes |n_{0{\bf r}}\rangle$ of the two spin bosons at a site ${\bf r}$, to a physical subspace $\mathcal{H}_{{\bf r}s}^{(0)}$ as follows
\begin{equation}
\mathcal{H}^{(0)}_{0{\bf r},{\bf r}} \rightarrow \mathcal{H}^{(0)}_{\bf r} \rightarrow \mathcal{H}_{{\bf r}s}^{(0)}.
\label{spinreduction}
\end{equation}
First, we restrict the Fock space $\mathcal{H}^{(0)}_{0{\bf r},{\bf r}}$ to a Fock space $\mathcal{H}^{(0)}_{\bf r}$ by excluding the $(0{\bf r})$-boson component, since its occupation $n_{0{\bf r}} = 2s-n_{\bf r}$ for any Fock state is determined by the $({\bf r})$-boson's occupation $n_{\bf r}$. Second, we limit an allowed number of excitations of the $({\bf r})$-boson in the physical Hilbert space $\mathcal{H}_{{\bf r}s}^{(0)}$ by an interval $0\leq n_{\bf r} \leq 2s$.
 
  The physics (dynamics, fluctuations, etc.) of each spin is determined by the creation and annihilation of the canonical-ensemble excitations (not bosons themselves!) in the system of two spin bosons via the operators 
\begin{equation}
\hat{\alpha}_{\bf r}^{'\dagger}=\theta(2s-\hat{n}_{\bf r})\hat{\alpha}_{\bf r}^{\dagger}, \qquad \hat{\alpha}^{'}_{\bf r}=\hat{\alpha}_{\bf r}\theta(2s-\hat{n}_{\bf r}), 
\label{spinalpha'}
\end{equation}
which leave invariant (i.e., do not lead out of) the physical subspace $\mathcal{H}_{{\bf r}s}^{(0)}$. They are the step-function $\theta(2s-\hat{n}_{\bf r})$ cutoff of the transition operators 
\begin{equation}
\hat{\alpha}_{\bf r}^{\dagger}=\hat{a}_{\bf r}^{\dagger}(1+2s-\hat{n}_{\bf r})^{-1/2}\hat{a}_{0{\bf r}}, \ \hat{\alpha}_{\bf r}=\hat{a}_{0{\bf r}}^{\dagger}(1+2s-\hat{n}_{\bf r})^{-1/2}\hat{a}_{\bf r}, 
\label{spinalpha}
\end{equation}
which are exactly similar to the Bogoliubov-Girardeau-Arnowitt ones in Eq. (\ref{alpha}).

    For every site ${\bf r}$, we introduce also a Fock space $\mathcal{H}_{\bf r}$, which spans the states $|n_{\bf r}\rangle$ of an unconstrained excitation at the site ${\bf r}$, and the creation and annihilation operators $\hat{\beta}_{\bf r}^{\dagger}$ and $\hat{\beta}_{\bf r}$ in that space, which obey a canonical commutation relation $[\hat{\beta}_{\bf r}, \hat{\beta}_{\bf r'}^{\dagger }]=\delta_{{\bf r},{\bf r'}}$. Their $\theta(2s-\hat{n}_{\bf r})$-cutoff counterparts are the subspace $\mathcal{H}_{{\bf r}s}$, which spans only the Fock states with a total occupation $n_{\bf r}\in [0,2s]$, and the operators
\begin{equation}
\hat{\beta}_{\bf r}^{'\dagger}=\theta(2s-\hat{n}_{\bf r})\hat{\beta}_{\bf r}^{\dagger}, \qquad \hat{\beta}^{'}_{\bf r}=\hat{\beta}_{\bf r}\theta(2s-\hat{n}_{\bf r}).
\label{spinhatbeta'}
\end{equation} 
They constitute a representation, which is exactly isomorphic to the original one in Eq. (\ref{spinalpha'}). It is valid in the total physical many-body Hilbert space of the system of $N$ spins $\mathcal{H}_s = \otimes_{\bf r} \mathcal{H}_{{\bf r}s}$, which is a tensor product of all physical Hilbert subspaces $\mathcal{H}_{{\bf r}s}$, and will be used from now on. In the sense of that isomorphism, one has
\begin{equation}
\hat{\alpha}_{\bf r}^{'}=\hat{\beta}_{\bf r}^{'}, \ \hat{\alpha}_{\bf r}^{'\dagger}=\hat{\beta}_{\bf r}^{'\dagger}, \ \hat{n}_{\bf r}=\hat{\alpha}_{\bf r}^{'\dagger}\hat{\alpha}_{\bf r}^{'} =\hat{\beta}_{\bf r}^{'\dagger}\hat{\beta}_{\bf r}^{'} \ \text{on} \ \mathcal{H}_s = \otimes_{\bf r} \mathcal{H}_{{\bf r}s}.
\label{spinalpha'=beta'}
\end{equation}

    The isomorphism in Eq. (\ref{spinalpha'=beta'}) is not trivial and is valid only on $\mathcal{H}_s$. First, the commutation relations for the creation and annihilation operators of the true spin excitations in Eqs. (\ref{spinalpha'}) and (\ref{spinhatbeta'}) are not canonical:
\begin{equation}
[\hat{\alpha}_{\bf r}^{'}, \hat{\alpha}_{\bf r'}^{'\dagger}]=\delta_{{\bf r},{\bf r'}}(1- (2s+1) \delta_{\hat{n}_{\bf r},2s}),
\label{spinnonCCRalpha}
\end{equation}
\begin{equation}
\qquad [\hat{\beta}_{\bf r}^{'}, \hat{\beta}_{\bf r'}^{'\dagger}]=\delta_{{\bf r},{\bf r'}}(1- (2s+1) \delta_{\hat{n}_{\bf r},2s}) .
\label{spinnonCCR}
\end{equation}
Second, it does not imply an isomorphism between the operators $\hat{\alpha}_{\bf r}, \hat{\alpha}_{\bf r}^{\dagger}$ on $\mathcal{H}^{(0)}_{\bf r}$ and the operators $\hat{\beta}_{\bf r}, \hat{\beta}_{\bf r}^{\dagger}$ on $\mathcal{H}_{\bf r}$, because the commutation relations for the latter are canonical, but the commutation relations for the former coincide with the noncanonical ones in Eq. (\ref{spinnonCCRalpha}).

   Finally, we present the explicit formulas for the vector components of the spin operator in Eqs. (\ref{spinvector})-(\ref{spin+-}) in terms of the creation and annihilation operators of the true spin excitations in Eq. (\ref{spinhatbeta'}):
\begin{equation}
\hat{S}^x_{\bf r}=\frac{1}{2} ({S}^{-}_{\bf r}+\hat{S}^{+}_{\bf r}), \ \hat{S}^y_{\bf r}=\frac{i}{2} ({S}^{-}_{\bf r}-\hat{S}^{+}_{\bf r}), \ \hat{S}^z_{\bf r}= s- \hat{n}_{\bf r},
\label{spinvectorbeta}
\end{equation}
\begin{equation}
\hat{S}^{+}_{\bf r} = \sqrt{2s-\hat{n}_{\bf r}} \hat{\beta}_{\bf r}^{'}, \ \hat{S}^{-}_{\bf r}= \hat{\beta}_{\bf r}^{'\dagger}\sqrt{2s-\hat{n}_{\bf r}}; \ \hat{n}_{\bf r}= \hat{\beta}_{\bf r}^{'\dagger}\hat{\beta}_{\bf r}^{'}.
\label{spin+-beta}
\end{equation} 

        A free Hamiltonian of a system of $N$ spins in a lattice
\begin{equation}
H_0 = \sum_{\bf r} \varepsilon \hat{n}_{\bf r}, \quad \hat{n}_{\bf r}=\hat{\beta}_{\bf r}^{\dagger}\hat{\beta}_{\bf r}, \quad \varepsilon= g \mu_B B_{ext},
\label{spinH_0}
\end{equation}
is determined by a Zeeman energy $-g\mu_B B_{ext}\hat{S}^z$ of a spin in an external magnetic field $B_{ext}$ (which is assumed homogeneous and directed along the axis $z$) via a $g$-factor and a Bohr magneton $\mu_B =e\hbar /(2Mc)$. We intentionally define the free Hamiltonian in Eq. (\ref{spinH_0}) via the unconstrained occupation operators $\hat{n}_{\bf r}=\hat{\beta}_{\bf r}^{\dagger}\hat{\beta}_{\bf r}$ on the extended many-body Hilbert space $\mathcal{H}= \otimes_{\bf r} \mathcal{H}_{\bf r}$ without any $\theta(2s-\hat{n}_{\bf r})$ cutoff factors. That makes the free Hamiltonian purely quadratic which is necessary for a validity of the standard diagram technique. The latter is crucial for a derivation of the Dyson-type equations, like Eqs. (\ref{spinbetaEq}) and (\ref{spinGEq}). One is allowed to skip the $\theta(2s-\hat{n}_{\bf r})$ cutoff factors in $H_0$ because of (i) an equality $\hat{\beta}_{\bf r}^{\dagger}\hat{\beta}_{\bf r}= \hat{\beta}_{\bf r}^{'\dagger}\hat{\beta}_{\bf r}^{'}$, which is valid on the physical many-body Hilbert space $\mathcal{H}_s$, and (ii) a fact that the occupation operator $\hat{n}_{\bf r}=\hat{\beta}_{\bf r}^{\dagger}\hat{\beta}_{\bf r}$ leaves invariant that space $\mathcal{H}_s$.
   
        However, the interaction Hamiltonian $H^{'}$ in the physical, constrained many-body Hilbert space $\mathcal{H}_s$ includes all terms with the nonpolynomial operator functions: the $\theta(2s-\hat{n}_{\bf r})$-cutoff factors and the functions $\sqrt{2s-\theta(2s-\hat{n}_{\bf r})\hat{n}_{\bf r}}$, coming from the $\hat{S}^x_{\bf r}$ and $\hat{S}^y_{\bf r}$ components of spins. As a generic example, we consider a well-known Heisenberg Hamiltonian \cite{Anderson1984,PatPokr,CritPhen-RG1992,Goldenfeld,Kadanoff,HolsteinPrimakoff,Dyson1956}
\begin{equation}
H^{'} = -\sum_{{\bf r}\neq {\bf r'}} J_{{\bf r},{\bf r'}}(\hat{S}_{\bf r}^x\hat{S}_{\bf r'}^x +\hat{S}_{\bf r}^y\hat{S}_{\bf r'}^y +\hat{S}_{\bf r}^z\hat{S}_{\bf r'}^z ),
\label{spinHint}
\end{equation}       
where a coupling between spins is usually assumed to be a symmetric function $J_{{\bf r},{\bf r'}}= J_{{\bf r}-{\bf r'}}$ of a vector ${\bf r}-{\bf r'}$, connecting spins. It contains the true spin excitations, Eq. (\ref{spinhatbeta'}), and, hence, should be written via the unconstrained excitation operators $\hat{\beta}^{\dagger}_{\bf r}$ and $\hat{\beta}_{\bf r}$ with an explicit account for all $\theta(2s-\hat{n}_{\bf r})$-cutoff factors and functions $\sqrt{2s-\theta(2s-\hat{n}_{\bf r})\hat{n}_{\bf r}}$, because one needs an exact account for all spin-constraint nonlinear interactions in the critical region. Thus, we find an exact result for the interaction Hamiltonian, presented here as a sum of two terms:
\begin{equation}
H^{'} = H^{'}_{xy} + H^{'}_z; 
\label{HxyHz}
\end{equation}
$$H^{'}_{xy}= -\sum_{{\bf r}\neq {\bf r'}} J_{{\bf r},{\bf r'}} \hat{S}_{\bf r}^{+}\hat{S}_{\bf r'}^{-}, \quad H^{'}_z = -\sum_{{\bf r}\neq {\bf r'}} J_{{\bf r},{\bf r'}}\hat{S}_{\bf r}^z\hat{S}_{\bf r'}^z .$$
The first one is related to the XY model, 
\begin{equation}
H^{'}_{xy} = -\sum_{{\bf r}\neq {\bf r'}} J_{{\bf r},{\bf r'}} \theta(2s-\hat{n}_{\bf r}) \theta(2s-1-\hat{n}_{\bf r'}) 
\label{Hxy}
\end{equation}
$$\times \sqrt{2s+1-\theta(2s+1-\hat{n}_{\bf r})\hat{n}_{\bf r}} \sqrt{2s-\theta(2s-\hat{n}_{\bf r'})\hat{n}_{\bf r'}} \hat{\beta}^{\dagger}_{\bf r} \hat{\beta}_{\bf r'} .$$
The second one is related to the Ising model,
\begin{equation}
H^{'}_z = -\sum_{{\bf r}\neq {\bf r'}} J_{{\bf r},{\bf r'}} [s-\theta (2s-\hat{n}_{\bf r})\hat{n}_{\bf r}] [s-\theta (2s-\hat{n}_{\bf r'})\hat{n}_{\bf r'}] .
\label{HIsing}
\end{equation}

    The result in Eqs. (\ref{HxyHz})-(\ref{HIsing}) agrees with the Holstein-Primakoff's one \cite{HolsteinPrimakoff}, but includes the additional $\theta(2s-\hat{n}_{\bf r})$-cutoff factors, which are very important for the rigorous microscopic theory of magnetic phase transitions in the critical region.  

   A total Hamiltonian $H=H_0 +H^{'}$ makes any operator $\hat{A}$, evolving in an imaginary time $\tau \in [0, \frac{1}{T}]$ in the Heisenberg representation, the Matsubara operator $\tilde{A}_{\tau}=e^{\tau H}\hat{A}e^{-\tau H}$. Let us introduce some notations. A symbol $\tilde{A}_{j\tau}$ stands for an operator itself $\tilde{A}_{1\tau} =\tilde{A}_{\tau}$ at $j=1$ and for a Matsubara-conjugated operator $\tilde{A}_{2\tau} =\tilde{\bar{A}}_{\tau}$ at $j=2$. A four-dimensional coordinate is $x=\{\tau,{\bf r}\}$. A product of all cutoff factors in a lattice of $N$ spins is 
\begin{equation}
\hat{\theta} =  \prod_{\bf r} \theta(2s-\hat{n}_{\bf r}).
\label{spintheta}
\end{equation}

   The quantities of primary interest in a microscopic theory of magnetic phase transitions are the mean value and correlations of the Matsubara annihilation and creation operators for spin excitations, both the unconstrained ($\tilde{\beta}_x, \tilde{\bar{\beta}}_x$) and true ($\tilde{\beta}^{'}_x = \tilde{\beta}_x \theta(2s-\tilde{n}_x), \tilde{\bar{\beta}}'_x =\theta(2s-\tilde{n}_x) \tilde{\bar{\beta}}_x$) ones. The definitions of the unconstrained (auxiliary) and true coherent order parameters and Matsubara Green's functions for spin excitations are as follows
\begin{equation}
\bar{\beta}_{\bf r}=\langle \hat{\beta}_{\bf r} \rangle , \quad \langle \dots \rangle \equiv  Tr \{\dots e^{-H/T}\}/Tr\{e^{-H/T}\} ,
\label{spinorder}
\end{equation}
\begin{equation}
\bar{\beta}^{'}_{\bf r}= \langle \hat{\beta}_{\bf r}^{'} \rangle_{\mathcal{H}_s}\equiv \frac{\langle \hat{\beta}_{\bf r}^{'}\hat{\theta} \rangle}{P_s} , \quad P_s = \langle \hat{\theta} \rangle ,
\label{spinorder'}
\end{equation}
\begin{equation}
G^{j_2\tau_2{\bf r_2}}_{j_1\tau_1{\bf r_1}}= -\langle T_{\tau} \tilde{\beta}_{j_1\tau_1{\bf r_1}}\tilde{\bar{\beta}}_{j_2\tau_2{\bf r_2}} \rangle ,
\label{spinGreen}
\end{equation}
\begin{equation}
G^{'j_2\tau_2{\bf r_2}}_{j_1\tau_1{\bf r_1}}= -\langle T_{\tau} \tilde{\beta'}_{j_1\tau_1{\bf r_1}}\tilde{\bar{\beta'}}_{j_2\tau_2{\bf r_2}} \hat{\theta} \rangle /P_s .
\label{spinGreen'}
\end{equation}   
Here $\langle \dots \rangle$ or $\langle \dots \rangle_{\mathcal{H}_s}$ mean the averages over the unconstrained or $\hat{\theta}$-cutoff many-body Hilbert spaces $\mathcal{H}$ or ${\mathcal{H}_s}$, respectively. They differ by the $\hat{\theta}$-cutoff factor from Eq. (\ref{spintheta}) and a normalization factor $P_s$, which is equal to a cumulative probability of all occupations of the spin excitations in the space $\mathcal{H}$ to be within the physically allowed intervals: $n_{\bf r} \in [0,2s]$ for all lattice sites ${\bf r}$.

\subsection{4.2. Exact equations for the order parameter and Green's functions of unconstrained spin excitations}

     As it was done in Sect. 3.2, we again use a remarkable fact that, after the first step of the many-body Hilbert space reduction in Eq. (\ref{spinreduction}) and due to the isomorphism in Eq. (\ref{spinalpha'=beta'}), it is possible to work in the unconstrained many-body Hilbert space $\mathcal{H}$, where the standard Matsubara (or Feynman) diagram technique and the exact Dyson equation are valid. Thus, we immediately get the exact microscopic equations for the unconstrained coherent order parameter and Green's functions of spin excitations via a total irreducible self-energy $\Sigma^{j_2x_2}_{j_1x_1}$ as follows
\begin{equation}
\bar{\beta}_{jx}= \check{G}^{(0)}[\check{\Sigma}[\bar{\beta}_{jx}]],
\label{spinbetaEq}
\end{equation}
\begin{equation}
G_{j_1x_1}^{j_2x_2}+ \bar{\beta}_{j_1x_1}\bar{\beta}^{*}_{j_2x_2} = G_{j_1x_1}^{(0)j_2x_2} + \check{G}^{(0)}[\check{\Sigma}[G_{j_1x_1}^{j_2x_2}]]. 
\label{spinGEq}
\end{equation}
Here the integral operators $\check{\Sigma}$ or $\check{G}^{(0)}$, applied to any function $f_{jx}$, stand for a convolution of that function $f_{jx}$ over the variables $j, \tau, {\bf r}$ with the total irreducible self-energy $\Sigma$ or a free propagator $G^{(0)}$, respectively:
\begin{equation}
\check{K}[f_{jx}]\equiv \sum_{j'=1}^2 \sum_{\bf r'} \int_0^{1/T} K^{j'x'}_{jx} f_{j'x'} d \tau' \quad \text{for} \ \check{K}=\check{\Sigma}, \check{G}^{(0)}.
\label{spinKEq}
\end{equation}
The free propagator is determined by exactly the same Eq. (\ref{G0}), that was used in Sect. 3.2, but with an appropriate for spins free Hamiltonian in Eq. (\ref{spinH_0}).
  
    The total irreducible self-energy is defined by equation
\begin{equation}
\langle T_{\tau} [\tilde{\beta}_{j_1x_1}, \tilde{H}^{'}_{\tau_1}]\Delta\tilde{\bar{\beta}}_{j_2x_2} \rangle =(-1)^{j_1}\sum_{j=1}^2 \int_0^{\frac{1}{T}} \sum_{\bf r}\Sigma_{j_1x_1}^{jx}G_{jx}^{j_2x_2} d\tau 
\label{spinself-energy}
\end{equation}
and exactly accounts for a possible coherent ordering ${\bar{\beta}}_{jx}$. We can substitute $\tilde{\beta}_{jx} = \bar{\beta}_{jx}+\Delta\tilde{\beta}_{jx}$ in the left-hand-side operator $\tilde{H}_{\tau_1}^{'}$ and use the nonpolynomial diagram technique \cite{PLA2015,KochLasPhys2007,KochJMO2007} together with the exact equations for the partial operator contractions, explained in Sections 3.3 and 4.3, or do the first-order, second-order or ladder approximation in interaction, when calculating the left-hand-side average in accord with the main theorem of the diagram technique in the interaction representation for the appropriate free and interaction Hamiltonians: 
\begin{equation}
H_0^{(\Delta)} = \sum_{\bf r} \varepsilon \Delta\hat{\beta}_{\bf r}^{\dagger}\Delta\hat{\beta}_{\bf r}, \quad H^{'(\Delta)} =H^{'} + H_0 - H_0^{(\Delta)}. 
\label{spinH_0^Delta}
\end{equation}
Of course, a final result does not depend on a way of splitting the total Hamiltonian in a particular pair of the free and interaction Hamiltonians. Both representations $H= H_0 +H'$ in Eqs. (\ref{spinH_0})-(\ref{spinHint}) and $H= H_0^{(\Delta)} +H^{'(\Delta)}$ in Eq. (\ref{spinH_0^Delta}) are equivalent and yield the same total irreducible self-energy (that is $\Sigma= \Sigma^{(\Delta)}$), if one takes into account the corresponding equivalent series of diagrams. A function in the left hand side of Eq. (\ref{spinGEq}) is known as an exact unconstrained connected Green's function $g_{j_1x_1}^{j_2x_2}$, defined in Eq. (\ref{g}). A strategy for calculation of the self-energy is similar to the BEC one (Sect. 3.2). Its full analysis is lengthy and will be discussed elsewhere.

\subsection{4.3. Fundamental equations for the constrained, physical spin excitations: The order parameter, Green's functions and partial operator contractions}   

     The true, constrained order parameter and Green's functions of spin excitations, Eqs. (\ref{spinorder'}) and (\ref{spinGreen'}), do not obey any standard closed equations of the Dyson type. A reason is that they are the averages which include, along with the usual products of the creation and annihilation operators, the nonpolynomial functions, like the ones ($\theta(2s-\hat{n}_{\bf r})$ or $\sqrt{2s-\hat{n}_{\bf r}}$) entering the true spin-excitation operators in Eq. (\ref{spinhatbeta'}) and the exact interaction Hamiltonian in Eq. (\ref{spinHint}). A standard diagram technique is not suited to deal with the nonpolynomial averages. 
    
    As in Sect. 3.3, we solve that problem by means of the nonpolynomial diagram technique \cite{PLA2015,KochLasPhys2007,KochJMO2007}. Namely, the true order parameter and Green's functions in Eqs. (\ref{spinorder'}) and (\ref{spinGreen'}) are given by the following explicit formulas
\begin{equation}
\bar{\beta}^{'}_{J}= \langle \tilde{s}_J[\tilde{\theta}_{\tau_i}] \rangle /P_s ,  \quad P_s = \langle \hat{\theta} \rangle ,
\label{spinbeta'}
\end{equation}
\begin{equation}
G^{'J_2}_{J_1}= -\langle \tilde{s}_{\bar{J}_2}[\tilde{s}_{J_1}[\tilde{\theta}_{\tau_1} \tilde{\theta}_{\tau_2}]] + \tilde{b}^{J_2}_{J_1}[\tilde{\theta}_{\tau_1} \tilde{\theta}_{\tau_2}] \rangle /P_s .
\label{spinG'}
\end{equation}
Here the basis partial one- and two-operator contractions, $\tilde{s}_J[f]$ and $\tilde{b}^{J_2}_{J_1}[f]$, are the operator-valued functionals, evaluated for an operator function $f$. We continue to use the short notations for the combined indexes, but this time in the coordinate, rather than momentum, representation that is more convenient for a lattice. Those indexes are $J= \{ji{\bf r_i}\}$ and $J_l= \{j_l i_l{\bf r_{i_l}}\}$ as well as $\bar{J}_l= \{\bar{j}_li_l{\bf r_{i_l}}\}$ with a complimentary index $\bar{j}_l=3-j_l$; $j_l=1,2$. An index $i=1,2$ (or $i_l$) enumerates different times $\tau_i$ (or $\tau_{i_l}$) in the external operator $\tilde{\beta}_{j\tau_i{\bf r_i}}$ (or $\tilde{\beta}_{j_l\tau_{i_l}{\bf r_{i_l}}}$). 

    A definition of the basis partial operator contractions is similar to the one discussed in Sections 3.3 and 3.4. Let us consider a generic case of an arbitrary operator function $f(\{\tilde{n}_{x_1},\tilde{n}_{x_2} \})$, which depends on the two sets $\{ \tilde{n}_{{\bf r_1}\tau_1} \}$ and $\{ \tilde{n}_{{\bf r_2}\tau_2} \}$ of the spin-excitation occupation operators at all lattice sites at two different times $\tau_1$ and $\tau_2$. The basis partial one-operator contraction
\begin{equation}
\tilde{s}_J[f(\{\tilde{n}_{x_1},\tilde{n}_{x_2} \})]\equiv \mathcal{A}_{\tau_i} T_{\tau}\{\tilde{\beta}_J^{c}f^c(\{\tilde{n}_{x_1},\tilde{n}_{x_2} \} ) \}
\label{spins}
\end{equation}
is defined by an explicit formula via the $f$'s Taylor series:
$$\tilde{s}_J[f]= T_{\tau} \sum_{\{m_{x_1}\}}^{\infty}\sum_{\{m_{x_2}\}}^{\infty} \sum_{\{m'_{x_1}\}}^{\{m_{x_1}\}}\sum_{\{m'_{x_1}\}}^{\{m_{x_2}\}} \frac{f^{(\{m_{x_1}\},\{m_{x_2}\})}}{\prod_{\{x_1,x_2\}} (m_{x_1}!m_{x_2}!)}$$
\begin{equation}
\times \prod_{\{x_1,x_2\}} \langle \mathcal{A}_{\tau_i} T_{\tau}\{\tilde{\beta}_J \tilde{n}_{x_1}^{m_{x_1}-m'_{x_1}}\tilde{n}_{x_2}^{m_{x_2}-m'_{x_2}}\} \rangle \tilde{n}_{x_1}^{m'_{x_1}}\tilde{n}_{x_2}^{m'_{x_2}};
\label{spinsTaylor}
\end{equation}
$$f^{(\{m_{x_1}\},\{m_{x_2}\})}=\frac{\partial^{\sum_{\{x_1\}} m_{x_1}+\sum_{\{x_2\}} m_{x_2}}f}{\prod_{\{x_1,x_2\}} \partial^{m_{x_1}}\tilde{n}_{x_1} \partial^{m_{x_2}}\tilde{n}_{x_2}} \Big|_{\{\tilde{n}_{x_l}=0\}}.$$ 
Here the sums run over all nonnegative integers $m_{x_i}$ and $m'_{x_i} \leq m_{x_i}$. A symbol $\mathcal{A}_{\tau_i}$ denotes an anti-normal ordering, which prescribes a position of the external operator $\tilde{\beta}_J$ relative to the function $f(\{\tilde{n}_{x_1},\tilde{n}_{x_2}\})$ and does not affect any other operators' positions, which are determined by the $T_{\tau}$-ordering. As is explained in Sect. 3.4, this definition implies a sum of all possible partial connected contractions, which were used in the nonpolynomial diagram technique in \cite{PLA2015,KochLasPhys2007,KochJMO2007} and denoted by the superscripts "c" in Eqs. (\ref{spins}), (\ref{spinb}). In the case, when the function $f$ depends only on a one-time set of spin-excitation occupations $\{ \tilde{n}_{x_1} \}$, the formulas for the basis partial operator contractions become much simpler. 

    The basis partial two-operator contraction 
\begin{equation}
\tilde{b}_{J_1}^{J_2}[f(\{\tilde{n}_{x_1},\tilde{n}_{x_2}\})] \equiv \mathcal{A}_{\tau_{i_1}\tau_{i_2}} T_{\tau} \{ \Delta \tilde{\beta}_{J_1}^{c}\Delta \tilde{\bar{\beta}}_{J_2}^{c} f^c(\{\tilde{n}_{\tau_1},\tilde{n}_{\tau_2}\}) \}
\label{spinb}
\end{equation}
is defined by a similar analytical formula
$$\tilde{b}_{J_1}^{J_2}[f]= T_{\tau} \sum_{\{m_{x_1}\}}^{\infty}\sum_{\{m_{x_2}\}}^{\infty} \sum_{\{m'_{x_1}\}}^{\{m_{x_1}\}}\sum_{\{m'_{x_1}\}}^{\{m_{x_2}\}} \frac{f^{(\{m_{x_1}\},\{m_{x_2}\})}}{\prod_{\{x_1,x_2\}} (m_{x_1}!m_{x_2}!)}$$
\begin{equation}
\prod_{\{x_1,x_2\}} \langle \mathcal{A}_{\tau_{i_1}\tau_{i_2}} T_{\tau}\{\Delta \tilde{\beta}_{J_1}\Delta \tilde{\bar{\beta}}_{J_2} \tilde{n}_{x_1}^{m_{x_1}-m'_{x_1}}\tilde{n}_{x_2}^{m_{x_2}-m'_{x_2}}\} \rangle \tilde{n}_{x_1}^{m'_{x_1}}\tilde{n}_{x_2}^{m'_{x_2}}.
\label{spinbTaylor}
\end{equation}
An anti-normal ordering $\mathcal{A}_{\tau_{i_1}\tau_{i_2}}$ prescribes only positions of the external operators $\Delta \tilde{\beta}_{J_1}$ and $\Delta \tilde{\bar{\beta}}_{J_2}$ relative to the function $f(\{\tilde{n}_{x_1},\tilde{n}_{x_2}\})$ and does not affect any other operators' positions, which are determined by the $T_{\tau}$-ordering.

    The exact closed difference (recurrence) equations for the basis partial operator contractions $\tilde{s}_{J}[f(\{ m_{J'} \})]$ and $\tilde{b}^{J_2}_{J_1}[f(\{ m_{J'} \})]$ for an arbitrary function $f(\{ m_{J'} \}) =f(\{\tilde{n}_{x_1}+2s+1-m_{x_1}, \tilde{n}_{x_2}+2s+1-m_{x_2}\})$, where a set $\{ m_{J'}\}$ consists of the two sets of integers $\{m_{x_1}\}$ and $\{m_{x_2}\}$, can be derived along the same lines as it was done for BEC in Sect. 3.4. Those equations have exactly the same universal form as Eqs. (\ref{1-contraction}) and (\ref{2-contraction}):
\begin{equation}
\tilde{s}_{J}[f(\{ m_{J'} \})]= \bar{\beta}_J f(\{ m_{J'} \})+ g_J^{J'} \Delta_{m_{J'}} \tilde{s}_{J'}[f(\{ m_{J'} \})],
\label{spin1-contraction}
\end{equation}
$$\tilde{b}^{J_2}_{J_1}[f(\{ m_{J'} \})] = -g_{J_1}^{J_2}f(\{ m_{J'} \}) - g_{J_1}^{J'} g_{J'}^{J_2} \Delta_{m_{J'}}f(\{ m_{J'} \})$$
\begin{equation}
+g_{J_1}^{J'_1} g_{J'_2}^{J_2} \Delta_{m_{J'_1}} \Delta_{m_{J'_2}} \tilde{b}^{J'_2}_{J'_1}[f(\{ m_{J'} \})].
\label{spin2-contraction}
\end{equation}

   A matrix $g_J^{J'}$, which enters these equations as the coefficients, is given by the solution of Eq. (\ref{spinGEq}) for the exact unconstrained connected Green's function in Eq. (\ref{g}). For a special case of the equal times $\tau_i=\tau_{i'}$, it is defined in accord with an anti-normal ordering of operators $\Delta \tilde{\beta}_J$, $\Delta \tilde{\bar{\beta}}_{J'}$. The latter is dictated by the anti-normal ordering in the definition of partial operator contractions in Eqs. (\ref{spins})-(\ref{spinbTaylor}). In Eqs. (\ref{spin1-contraction})-(\ref{spin2-contraction}), a symbol $\Delta_{m_{J'}}$ means a partial difference operator \cite{DE-Elaydi}, defined after Eq. (\ref{2-contraction}), and we assume an Einstein's summation over the repeated indexes $J', J'_1, J'_2$. The sums run over $j'=1,2$ and all different arguments $\tilde{n}_{x'_{i'}}$ of the function $f$, enumerated by $i'=1,2$ and ${\bf r'_{i'}}$, for $J'$ and similarly for $J'_1, J'_2$. 

    The equations (\ref{spin1-contraction}) and (\ref{spin2-contraction}) are the linear systems of the integral equations over the spin positions' variables and the discrete (recurrence) equations over the variables $\{m_{J'}\}$ with the well-known methods of solution \cite{PDE-Cheng,DE-Agarwal,DE-Elaydi}, e.g., via a Z-transform (a discrete analog of a Laplace transform), a characteristic function (a solution for $f=\exp(\sum_{x_i} iu_{x_i}\tilde{n}_{x_i})$ with a subsequent Fourier transform), or a direct recursion. The operator contractions in Eqs. (\ref{spinbeta'})-(\ref{spinG'}) are given by those solutions at $m_{J'}=2s+1$. 

    Finally, their average in Eqs. (\ref{spinbeta'})-(\ref{spinG'}) amounts to the averages like $\langle f(\{\tilde{n}_{x_1},\tilde{n}_{x_2}\}) \rangle$, that is reduced to a calculation of a joint probability distribution $\rho_{\{n_{x_1},n_{x_2}\}}$ of the noncommuting operators $\tilde{n}_{x_1}$ and $\tilde{n}_{x_2}$. The latter can be done similar to \cite{PRA2010,PRA2014} by the method of the characteristic function, mentioned in Sect. 3.3 (see an example in Sect. 5.2, 5.4). The macroscopic wave function (the coherent order parameter) of the spin excitations' coherence $\bar{\beta}_{\bf r}\neq 0$, entering Eqs. (\ref{spinGEq}), (\ref{spinbeta'}) and (\ref{spin1-contraction}), makes the quasiparticles stable and can be found from Eq. (\ref{spinbetaEq}).

\section{5. On the exact solution for the three-dimensional Ising model}

     The exact results for the constrained Green's functions, the order parameter and the partial operator contractions of the presented above microscopic theory provide the regular, canonical tools for finding the exact solutions for the particular many-body Hamiltonians, including the 3D systems. This is in contrast to the known exactly solvable models, the solutions for which are based on some special mathematical tricks, applicable only due to very specific degenerate properties of those models and their low (1D or 2D) dimensionality (see, for example, \cite{Baxter1989}). Typically, those models, for example, the 2D Ising models, demonstrate a critical behavior, which is very different from the behavior of the actual 3D systems. As a generic example, we discuss here the exact solution for the 3D Ising model, finding of which is a long-standing problem.

    Let us consider $N$ spins $s=1/2$ in a lattice, described in Sect. 4.1, with the free Hamiltonian $H_0$ in Eq. (\ref{spinH_0}) and the interaction Hamiltonian $H'_z$ in Eq. (\ref{HIsing}) in the absence of the term $H'_{xy}$ in Eq. (\ref{HxyHz}) and zero coherent ordering ${\bar{\beta}}_{\bf r}=0$. For a spin at a site ${\bf r_0}$ there is the coordination number $p$ of the neighboring spins at sites ${\bf r_l}= {\bf r_0} + {\bf l}, l=1, ..., p,$ which have the nonzero couplings $J_{{\bf r_0},{\bf r_l}} \neq 0$. The coupling $J_{{\bf r_0},{\bf r}}$ between the spin ${\bf r_0}$ and any other spin ${\bf r} \neq {\bf r_l}$ vanishes. We assume an arbitrary dimensionality of the lattice $d=1, 2, 3, ...$. A quantity of primary interest is the equal-time anti-normally ordered correlation matrix $g^{I'}_I$, which is equal to the unconstrained connected Green's function $G^{J'}_J + \bar{\beta}_J \bar{\beta}^{*}_{J'}$ at $\tau' \to \tau -(-1)^{j'}\times 0$. From now on we use the combined indexes $I=\{j,{\bf r}\}, I'=\{j',{\bf r'}\}, I_0=\{j_0,{\bf r_0}\}$, etc.

\subsection{5.1. Exact solution for a self-energy}

    The first, main step is finding an exact explicit formula for the total irreducible self-energy from Eq. (\ref{spinself-energy}):
\begin{equation}
\Sigma_{J_0}^{J} =-\delta(\tau-\tau_0) \sum_{l=1}^{p} \sum_{I'} J_{{\bf r_0},{\bf r_l}} \bar{b}_{I_0}^{I'}[f(\tilde{n}_{\bf r_0}-1,\tilde{n}_{\bf r_l})] (g^{-1})_{I'}^{I} . 
\label{IsingSelfEnergy}
\end{equation}
Here $\bar{b}_{I_0}^{I'}[f]=\langle\tilde{b}_{I_0}^{I'}[f]\rangle$ is an unconstrained average of the basis partial two-operator contraction in Eq. (\ref{spinb}),
\begin{equation}
f(\tilde{n}_{\bf r_0},\tilde{n}_{\bf r_l}) =(\delta_{0,\tilde{n}_{\bf r_0}}-\delta_{1,\tilde{n}_{\bf r_0}}) (1-2\delta_{1,\tilde{n}_{\bf r_l}}),
\label{fIsing}
\end{equation}
and $g^{-1}$ denotes the matrix, which is inverse to $g$. The exact recurrence Eq. (\ref{spin2-contraction}) immediately proves that for a given spin ${\bf r_0}$ the self-energy matrix in Eq. (\ref{IsingSelfEnergy}) is not zero only for ${
\bf r} ={\bf r_l}, l=0, 1, ...,p$. In other words, it has a pure $(p+1)$-banded diagonal structure in indexes ${\bf r_0}, {\bf r}$:
\begin{equation}
\Sigma_{J_0}^{J} =\delta(\tau-\tau_0) \sum_{l=0}^{p} \delta_{{\bf r},{\bf r_l}} \Sigma_{j_0{\bf r_0}}^{j{\bf r_l}}(l) , \quad {\bf r_l}= {\bf r_0} + {\bf l}.
\label{p+1SelfEnergyIsing}
\end{equation}
Moreover, we can easily and exactly find each $2\times2$-matrix block $\Sigma_{j_0}^{j}(l)$ by solving the recurrence Eq. (\ref{spin2-contraction}). That result is crucial for the exact solution of the Ising model. 
 
    For simplicity's sake, we consider below only a particular, usually discussed case of the homogeneous phases, when all sites have the same auto-correlation matrices and other properties. In this case the order parameter is homogeneous along the lattice and the Green's function $G^{j_2\tau_2{\bf r_2}}_{j_1\tau_1{\bf r_1}}$ does not depend separately on each of the two position vectors ${\bf r_1}$ and ${\bf r_2}$, but depends only on their difference ${\bf r_2}-{\bf r_1}$. So, it is a Toeplitz matrix with respect to the indexes ${\bf r_1}$ and ${\bf r_2}$. A generalization to an arbitrary case will be presented elsewhere.
 
    The corresponding $2\times2$-matrices $\bar{b}_{j_0}^{j'}(l) =\bar{b}_{j_0{\bf r_0}}^{j'{\bf r_l}}[f(\tilde{n}_{\bf r_0}-1,\tilde{n}_{\bf r_l})]$ over the indexes $j_0$ and $j'$ are found as follows 
$$\bar{b}(l=0)= 2\rho_{1,1}-\rho_1 +2\rho_{0,1}K -\rho_0 S^{-1} ,$$
\begin{equation}
\bar{b}(l\neq 0)= 2\rho_{0,0} KQCH -2\rho_{1,0} KCQ .
\label{bl}
\end{equation}  

Here, for a given pair of neighboring sites $\{{\bf r_0}, {\bf r_l}\}$, we work with the correlation hermitian $4\times4$-matrix $q=q^{\dagger}$,
\begin{equation} 
q_I^{I'}(l) \equiv g_{j{\bf R}}^{j'{\bf R'}} =-\langle \mathcal{A} \hat{\beta}_{j{\bf R}} \hat{\beta}^{\dagger}_{j'{\bf R'}} \rangle , \ {\bf R}, {\bf R'} \in \{{\bf r_0}, {\bf r_l} \},
\label{q4x4}
\end{equation}
where the lattice-site indexes ${\bf R}$ and ${\bf R'}$ run only over two values $\{ {\bf r_0}, {\bf r_l} \}$ and $\mathcal{A}$ means the anti-normal ordering. We introduce the correlation $2\times2$-matrix $g(l)$ as well as the shortened notations for its auto-correlation, $S=g(0)=S^{\dagger}$, and 
cross-correlation, $C=g(l\neq 0)$, versions:
\begin{equation}
g_{j}^{j'}(l)= g_{j{\bf r_0}}^{j'{\bf r_l}}, \quad S_{j}^{j'}= g_{j{\bf r_0}}^{j'{\bf r_0}}, \quad C_{j}^{j'}= g_{j{\bf r_0}}^{j'{\bf r_l}} (l\neq 0) .
\label{SandC}
\end{equation}
Deriving Eq. (\ref{bl}), we introduce the $2\times2$-matrices 
\begin{equation}
Q=S^{-1}, \ K=(S-CS^{-1}C^{\dagger})^{-1}, \ H=(S-C^{\dagger}S^{-1}C)^{-1},
\label{QKH}
\end{equation}
needed for the inversion of the block matrix $q$ in Eq. (\ref{q4x4}),
\begin{equation}
q= \Big( \frac{S \ \ | \ C}{C^{\dagger}| \ S} \Big) =q^{\dagger} ,
\label{q-block}
\end{equation}
by means of the well-known Frobenius formulas. 

    The result in Eq. (\ref{bl}) includes the single-site and two-sites, joined uncutoff probability distributions,
\begin{equation}
\rho_{n_{\bf r_0}}=\langle \delta_{\tilde{n}_{\bf r_0}, n_{\bf r_0}} \rangle, \quad \rho_{n_{\bf r_0},n_{\bf r_l}}=\langle \delta_{\tilde{n}_{\bf r_0}, n_{\bf r_0}} \delta_{\tilde{n}_{\bf r_l}, n_{\bf r_l}} \rangle ,
\label{rho01}
\end{equation}
for the spin-boson occupation operators $\tilde{n}_{\bf r_0}$ and $\tilde{n}_{\bf r_l}$ at the neighboring sites ${\bf r_0}$ and ${\bf r_l}$ to acquire the $n_{\bf r_0}$ and $n_{\bf r_l}$ quanta of excitations, respectively. In Sect. 5.2 we calculate those occupation distributions exactly as well.

   The self-energy blocks $\Sigma (l)$ in Eq. (\ref{p+1SelfEnergyIsing}) for neighboring sites ${\bf r_l}, l=0,1,...,p$, can be calculated via a product of two $4\times4$-matrices, in which the index ${\bf R'}$ in $I'=\{ j',{\bf R'} \}$ runs only over the two sites $\{ {\bf r_0},{\bf r_l} \}$, as follows
\begin{equation}
\Sigma_{j_0{\bf r_0}}^{j{\bf r_0}}(0) =-\sum_{l=1}^{p}\sum_{I'}J_{{\bf r_0},{\bf r_l}} \bar{b}_{j_0{\bf r_0}}^{j'{\bf R'}}[f(\tilde{n}_{\bf r_0}-1,\tilde{n}_{\bf r_l})] (q^{-1}(l))_{j'{\bf R'}}^{j{\bf r_0}},  
\label{Sigma4x4}
\end{equation}
$$\Sigma_{j_0{\bf r_0}}^{j{\bf r_l}}(l\neq 0) =-\sum_{I'}J_{{\bf r_0},{\bf r_l}} \bar{b}_{j_0{\bf r_0}}^{j'{\bf R'}}[f(\tilde{n}_{\bf r_0}-1,\tilde{n}_{\bf r_l})] (q^{-1}(l))_{j'{\bf R'}}^{j{\bf r_l}}.$$    
A result for the $2\times2$-matrix blocks $\Sigma(l)=(\Sigma_{j_0{\bf r_0}}^{j{\bf r_l}}(l))$ is
$$\Sigma(0)= -\sum_{l=1}^{p} J_{{\bf r_0},{\bf r_l}} [2\rho_{1,1}K +2\rho_{0,1}K^2 -\rho_1 S^{-1} -\rho_0 S^{-2}$$ 
\begin{equation}
+2\rho_{1,0}KCQ^{2}C^{\dagger}K+2\rho_{0,0}KQCHQC^{\dagger}K],
\label{Sigma(0)}
\end{equation}  
$$\Sigma(l \neq 0)= J_{{\bf r_0},{\bf r_l}} \{2\rho_{1,1}KCQ +2\rho_{0,1}K^{2}CQ$$ 
\begin{equation}
+[2\rho_{1,0} KCQ^{2}+2\rho_{0,0}KQCHQ](1+C^{\dagger}KCQ)\}.
\label{Sigma(l)}
\end{equation}

\subsection{5.2. Exact solution for the single-site and two-sites, joined statistics of the spin-boson occupations}

   The second step is finding the distributions in Eq. (\ref{rho01}). The single-site distribution of the spin-boson excitations
\begin{equation}
\rho_n \equiv \langle \delta_{\tilde{n}_{\bf r},n} \rangle = \frac{\langle \tilde{b}_{1{\bf r}}^{1{\bf r}}[\delta_{\tilde{n}_{\bf r}-1,n}] \rangle}{n+1}, \ n=0,1,2, \dots ,
\label{rho-b}
\end{equation}
can be calculated via the partial two-operator contraction $\tilde{b}_{1{\bf r}}^{1{\bf r}} [\delta_{\tilde{n}_{\bf r}-1, n}]$ again by means of the recurrence Eq. (\ref{spin2-contraction}), which yields the following recurrence equation for $\rho_n$:
\begin{equation}
\rho_{n+2} -\frac{2n+3}{n+2}(Q_1^1 +1)\rho_{n+1} +\frac{n+1}{n+2}[(Q_1^1+1)^2 -|Q_1^2|^2]\rho_n =0;
\label{rho-recurrence}
\end{equation} 
\begin{equation}
Q_1^1=\frac{g_{1{\bf r}}^{1{\bf r}}}{(g_{1{\bf r}}^{1{\bf r}})^2 -|g_{1{\bf r}}^{2{\bf r}}|^2}, \quad Q_1^2=-\frac{g_{1{\bf r}}^{2{\bf r}}}{(g_{1{\bf r}}^{1{\bf r}})^2 -|g_{1{\bf r}}^{2{\bf r}}|^2} .
\label{Q11Q12}
\end{equation}
We find its solution via the well-known Jacobi polynomial $P_n^{(\alpha,\beta)}$ or hypergeometric function, defined by a generalized hypergeometric series $F(a,b;c;z)=\sum_{k=0}^{\infty} \frac{(a)_k(b)_kz^k}{(c)_k k!}$:
\begin{equation}
\rho_n =\frac{\rho_0}{\zeta_1^n}P_n^{(0,-n-\frac{1}{2})} (\frac{2\zeta_1}{\zeta_2}-1) =\frac{\rho_0}{\zeta_2^n}F(-n,\frac{1}{2};1;\frac{\zeta_1-\zeta_2}{\zeta_1});
\label{rhoJacobi}
\end{equation}
\begin{equation}
\rho_0 = \Big[ \frac{(1-\zeta_1)(1-\zeta_2)}{\zeta_1 \zeta_2} \Big]^{\frac{1}{2}}, \ \zeta_{1,2} =\frac{g_{1{\bf r}}^{1{\bf r}}\mp |g_{1{\bf r}}^{2{\bf r}}|}{1+g_{1{\bf r}}^{1{\bf r}} \mp |g_{1{\bf r}}^{2{\bf r}}|} .
\label{rho0}
\end{equation}

    Another way to find the same $\rho_n$ is to use its relation
\begin{equation}
\rho_n =\int_{-\pi}^{\pi}\frac{\theta(u)}{2\pi}e^{-iu}du =\frac{1}{n!}\frac{\partial^n\theta}{\partial z^n}\Big|_{z=0}, \ z=e^{iu},
\label{rhoTheta}
\end{equation}
to a characteristic function $\theta(u)$, obeying an equation
\begin{equation}
z\frac{d\theta}{dz} =-\theta +\langle \tilde{b}_{1{\bf r}}^{1{\bf r}}[e^{iu(\tilde{n}_{\bf r}-1)}] \rangle, \quad \theta(u) \equiv \langle e^{iu\tilde{n}_{\bf r}} \rangle . 
\label{Theta}
\end{equation}
In case of a function $f=e^{iu\tilde{n}_{\bf r}}$, the recurrence Eq. (\ref{spin2-contraction}) for the $2\times2$-matrix of the two-operator contraction $\hat{b}=(\tilde{b}_{j_1\tau{\bf r}}^{j_2\tau{\bf r}}[e^{iu\tilde{n}_{\bf r}}])$ becomes a simple $2\times2$-matrix equation
\begin{equation}
\hat{b}=\Big( \frac{1-z}{z}\Big)^2S\hat{b}S-\Big( 1+\frac{1-z}{z}S\Big) Se^{iu\tilde{n}_{\bf r}}.
\label{b2x2}
\end{equation}
We find its solution, plug it into Eq. (\ref{Theta}), and solve that equation. A result for the characteristic function is
\begin{equation}
\theta(u) =\Big[ \frac{(1-\zeta_1)(1-\zeta_2)}{(z-\zeta_1)(z-\zeta_2)} \Big]^{\frac{1}{2}} \equiv \frac{\rho_0 \sqrt{\zeta_1\zeta_2}}{\sqrt{(z-\zeta_1)(z-\zeta_2)}} .
\label{ThetaFinal}
\end{equation}
Plugging it into Eq. (\ref{rhoTheta}), we obtain the same distribution (\ref{rhoJacobi}) that was found above by a completely different method of Eq. (\ref{rho-b}). In particular, we find the probability 
\begin{equation}
\rho_1 = \rho_0 \Big( 1+ \frac{g_{1{\bf r}}^{1{\bf r}}}{(g_{1{\bf r}}^{1{\bf r}})^2 -|g_{1{\bf r}}^{2{\bf r}}|^2} \Big) ,
\label{rho1}
\end{equation}
requested by Eqs. (\ref{Sigma(0)}) and (\ref{Sigma(l)}) for the self-energy. 

    Thus, the problem of finding the single-site probability distribution of the spin boson excitations via the auto-correlation coefficients $g_{1{\bf r}}^{1{\bf r}}$ and $g_{1{\bf r}}^{2{\bf r}}$ is fully solved.

    A solution of a similar problem for the joined distribution of spin excitations at two sites $\{{\bf r_1},{\bf r_2}\}$ in Eq. (\ref{rho01}), 
\begin{equation}
\rho_{n_1,n_2}= \frac{1}{n_1!n_2!}\frac{\partial^{n_1+n_2}\theta}{\partial z_1^{n_1}\partial z_2^{n_2}}\Big|_{z_1=z_2=0}, \ z_{\bf r_k}\equiv z_k=e^{iu_k},
\label{rhoTheta2}
\end{equation}
via the two equations for the characteristic function
 
$\qquad \qquad \theta(u_1,u_2)=\langle e^{iu_1\tilde{n}_{\bf r_1}+iu_2\tilde{n}_{\bf r_2}}\rangle,$
\label{Theta2}
\begin{equation}
z_k\frac{\partial \theta}{\partial z_k} =-\theta +\langle \tilde{b}_{1{\bf r_k}}^{1{\bf r_k}}[e^{iu_1\tilde{n}_{\bf r_1}+iu_2\tilde{n}_{\bf r_2}-iu_k}] \rangle , k=1,2,
\label{Theta2}
\end{equation}
can be done in the same way. In that case of a function $f=e^{iu_1\tilde{n}_{\bf r_1}+iu_2\tilde{n}_{\bf r_2}}$, the recurrence Eq. (\ref{spin2-contraction}) for the $4\times4$-matrix of the two-operator contraction $\hat{b}=(\tilde{b}_{j\tau{\bf r}}^{j'\tau{\bf r'}}[e^{iu_1\tilde{n}_{\bf r_1}+iu_2\tilde{n}_{\bf r_2}}])$, where $j,j' \in \{ 1,2 \}$ and ${\bf r},{\bf r'} \in \{ {\bf r_1},{\bf r_2} \}$ , becomes a more complex $4\times4$-matrix equation
\begin{equation} 
Z\frac{1}{q}\hat{b} -\hat{b}\frac{1}{Z}q = (q-Z)f; \ Z=\text{diag} \{ \xi_I \}, \xi_{j{\bf r_k}}\equiv \xi_k=\frac{z_k}{z_k -1}.
\label{b4x4}
\end{equation}
Here the $4\times4$-matrix $q$ is defined in Eq. (\ref{q-block}) by the $2\times2$-matrices $S$ and $C$ of the auto- and cross-correlations for two sites $\{{\bf r_1},{\bf r_2}\}$. A matrix Eq.(\ref{b4x4}) has a well-known in algebra type and can be solved by the regular methods. Moreover, in our case we find a simple explicit solution
\begin{equation} 
\hat{b} =-(1+qZ^{-1})^{-1} q e^{iu_1\tilde{n}_{\bf r_1}+iu_2\tilde{n}_{\bf r_2}}.
\label{b4x4solution}
\end{equation}
Plugging it into Eqs. (\ref{Theta2}) and solving them, we find an explicit exact solution for the characteristic function
\begin{equation}
\theta(u_1,u_2)= \frac{1}{(z_{\bf r_1}-1)(z_{\bf r_2}-1) \sqrt{Q(\xi_1,\xi_2)}},
\label{Theta2Final}
\end{equation}
where we use the shortened notations for the basis normal and anomalous auto- and cross-correlations
\begin{equation}
g_{1{\bf r_k}}^{1{\bf r_k}}=g_{11},\ g_{1{\bf r_k}}^{2{\bf r_k}}=g_{12}, \ g_{1{\bf r_1}}^{1{\bf r_2}}=c_{11}, \ g_{1{\bf r_1}}^{2{\bf r_2}}=c_{12},
\label{g12c12}
\end{equation}
and a quadratic, in each variable $\xi_{1,2}$, polynomial, defined by a determinant of a following $4\times4$-matrix,
\begin{equation}
Q(\xi_1,\xi_2)= \text{det} (q+Z) \equiv \text{det} \Big( \frac{S+\xi_1 | \quad \ C \ }{ \quad C^{\dagger} \ \ \ |S+\xi_2} \Big) .
\label{Q(1,2)}
\end{equation}

   This result yields the joined probabilities of two neighboring sites occupations $\rho_{n_1,n_2}$, entering the self-energy in Eqs. (\ref{Sigma(0)})-(\ref{Sigma(l)}), as the simple derivatives in Eq. (\ref{rhoTheta2}) of an elementary function in Eq. (\ref{Theta2Final}). For simplicity's sake, we present them for the case, when $g_{1{\bf r_k}}^{2{\bf r_k}}=g_{1{\bf r_k}}^{2{\bf r_k*}}\geq 0$ and the cross-correlation between neighboring sites is symmetric: $g_{1{\bf r_1}}^{1{\bf r_2}}=g_{1{\bf r_1}}^{1{\bf r_2*}}, \rho_{1,0}=\rho_{0,1}$. The former can be always achieved by adjusting phases of the spin boson creation operators. The latter implies a symmetry:
$$Q(\xi_1,\xi_2)= [(g_{11}+\xi_1)^2 -g_{12}^2][(g_{11}+\xi_2)^2 -g_{12}^2]-2(g_{11}+\xi_1)\times$$ 
$(g_{11}+\xi_2)(c_{11}^2+|c_{12}^2|) +2(2g_{11}+\xi_1+\xi_2)c_{11}g_{12} (c_{12}+c_{12}^*) -$
\begin{equation}
g_{12}^2(c_{12}^2+c_{12}^{*2}) -2c_{11}^2 g_{12}^2 +(c_{11}^2 -|c_{12}^2|)^2 .
\label{Q(1,2)sym}
\end{equation}   
Then, a direct differentiation yields the probabilities
\begin{equation}
\rho_{0,0}= \frac{1}{\sqrt{\text{det} q}},
\label{rho00}
\end{equation}
\begin{equation}
\rho_{1,0}=\rho_{0,0}\Big[ 1+ (q^{-1})_{1{\bf r_1}}^{1{\bf r_1}} \Big] , 
\label{rho10}
\end{equation}
\begin{equation}
\rho_{1,1}= \rho_{0,0} \{ [ 1+ (q^{-1})_{1{\bf r_1}}^{1{\bf r_1}}] [ 1+ (q^{-1})_{1{\bf r_2}}^{1{\bf r_2}}]
\label{rho11}
\end{equation}
$$+ (q^{-1})_{1{\bf r_1}}^{1{\bf r_2}}(q^{-1})_{1{\bf r_2}}^{1{\bf r_1}} + (q^{-1})_{1{\bf r_1}}^{2{\bf r_2}}(q^{-1})_{2{\bf r_2}}^{1{\bf r_1}} \}.$$

    Thus, we find the two-sites, joined probability distribution of the spin boson excitations via the basis normal and anomalous auto- and cross-correlations in Eq. (\ref{g12c12}).

\subsection{5.3. The exact consistency equations: Solution for the basis auto- and cross-correlations}

    So, the total irreducible self-energy and the spin-boson occupation statistics are known exactly via $(1+p)$ basis correlation $2\times2$-matrices $g(l), l=0,1,...,p$, in Eq. (\ref{SandC}) alone. Actually, as is shown below, the Green's functions, the order parameter and other statistical and thermodynamic quantities are also determined by those auto-correlations $g(0)=(g_{j{\bf r_0}}^{j'{\bf r_0}})$ for a spin boson at a site ${\bf r_0}$ and its cross-correlations $g(l\ne 0)=(g_{j{\bf r_0}}^{j'{\bf r_l}})$ with the coordination number $p$ of the neighboring spins at sites ${\bf r_l}={\bf r_0}+{\bf l}$. Due to the complex-conjugation relations
\begin{equation}
g_{1{\bf r_0}}^{1{\bf r_0}}=g_{2{\bf r_0}}^{2{\bf r_0}}, \ g_{1{\bf r_0}}^{2{\bf r_0}}=g_{2{\bf r_0}}^{1{\bf r_0}*}, \ g_{1{\bf r_0}}^{1{\bf r_l}}=g_{2{\bf r_0}}^{2{\bf r_l}*}, \ g_{1{\bf r_0}}^{2{\bf r_l}}=g_{2{\bf r_0}}^{1{\bf r_l}*},
\label{cc-relations}
\end{equation}
there are only two independent, normal $g_{1{\bf r_0}}^{1{\bf r_l}}$ and anomalous $g_{1{\bf r_0}}^{2{\bf r_l}}$, correlation parameters per each basis correlation $2\times2$-matrix, that is, only $2(1+p)$ numbers, which determine the characteristics of critical phenomena. 

    Therefore, the very crucial, third step is to find the exact consistency equations for those $2(1+p)$ basis auto- and cross-correlations. We solve this problem in two steps. First, we solve the Dyson-type Eq. (\ref{spinGEq}) for the unconstrained Green's functions in terms of those basis correlations. Second, we close the loop by expressing the basis correlations themselves via those Green's functions. 

    For the considered stationary homogeneous phases, the Green's functions, the equal-time correlation functions, and the self-energy depend only on the differences of their arguments $\tau= \tau_1-\tau_2$ and ${\bf r}= {\bf r_2}-{\bf r_1}$, that is
\begin{equation}
G_{J_1}^{J_2}=G_{j_1j_2}(\tau, {\bf r}), \ g_{j_1{\bf r_1}}^{j_2{\bf r_2}}=g_{j_1j_2}({\bf r}), \ \Sigma_{J_1}^{J_2}=\delta(\tau)\Sigma_{j_1j_2}({\bf r}). 
\label{r-r}
\end{equation}
Hence, it is straightforward to solve the Dyson-type Eq. (\ref{spinGEq}) by means of the Fourier transformation over imaginary time $\tau \in [-\frac{1}{T},\frac{1}{T}]$ and the discrete Fourier transformation over space. The latter has a following form
\begin{equation}
g({\bf k})=\sum_{\bf r}g({\bf r})e^{-i{\bf k}{\bf r}}, \ g({\bf r})= \Big( \frac{a}{L} \Big)^d \sum_{\bf k}g({\bf k})e^{i{\bf k}{\bf r}}, 
\label{DFT}
\end{equation}
where the sums run over all lattice sites  ${\bf r}$ with a period $a$ and discrete wave vectors ${\bf k}=\{ k_i; i=x,y,z \}, k_i= \frac{2\pi}{L}q$ with an integer $q$; $k_i \in [-\frac{\pi}{a},\frac{\pi}{a}]$. We discern the Fourier transform and its inverse by the arguments ${\bf k}$ and ${\bf r}$. The result for the normal and anomalous Green's functions is
\begin{equation}
G_{11}(\tau,{\bf k})= \sum_{j=1}^{2} (-1)^j \frac{[i\omega^{(j)}+\varepsilon+\Sigma_{22}({\bf k})]e^{i\omega^{(j)}(\frac{\text{sign}(\tau)}{2T}-\tau)}}{2(\omega^{(2)}-\omega^{(1)})\sin [\omega^{(j)}/(2T)]} ,
\label{G11}
\end{equation}
\begin{equation}
G_{12}(\tau,{\bf k})= \sum_{j=1}^{2} \frac{(-1)^j \Sigma_{12}({\bf k})e^{i\omega^{(j)}(\frac{\text{sign}(\tau)}{2T}-\tau)}}{2(\omega^{(1)}-\omega^{(2)})\sin [\omega^{(j)}/(2T)]} ,
\label{G12}
\end{equation}
where the two quasiparticle eigen-energies 
\begin{equation}
i\omega^{(1,2)}= \frac{\Sigma_{11}-\Sigma_{22}}{2} \pm \Big[ \Big( \varepsilon +\frac{\Sigma_{11}+\Sigma_{22}}{2} \Big)^2 -\Sigma_{12}\Sigma_{21} \Big]^{\frac{1}{2}} 
\label{omega12}
\end{equation}
depend on the wave vector ${\bf k}$ via the self-energies 
\begin{equation}
\Sigma_{j_0j}({\bf k})= \sum_{l=0}^p \Sigma_{j_0}^{j}(l) e^{-i{\bf k}({\bf r_l}-{\bf r_0})} . 
\label{Sigma(k)}
\end{equation}
The $2\times2$-matrix blocks $\Sigma(l)$ are found in Eqs. (\ref{Sigma4x4})-(\ref{Sigma(l)}).

    The spatial Fourier transforms of the normal and anomalous equal-time correlation functions follow from Eqs. (\ref{G11}) and (\ref{G12}) in the limit $\tau \to +0$:
\begin{equation} 
g_{11}({\bf k})= \sum_{j=1}^{2} \frac{(-1)^j [i\omega^{(j)}+\varepsilon+\Sigma_{22}({\bf k})]}{i(\omega^{(1)}-\omega^{(2)})[1-\exp (-i\omega^{(j)}/T)]} ,
\label{g11}
\end{equation}
\begin{equation} 
g_{12}({\bf k})= \sum_{j=1}^{2} \frac{(-1)^j \Sigma_{12}({\bf k})}{i(\omega^{(2)}-\omega^{(1)})[1-\exp (-i\omega^{(j)}/T)]} .
\label{g12}
\end{equation}

    Thus, we derive the equations for the values of normal and anomalous correlation functions at $(1+p)$ difference position vectors ${\bf l}= {\bf r_l}-{\bf r_0}$ of the neighboring spins:
\begin{equation} 
g_{1j}(l)= \Big( \frac{a}{L} \Big)^d \sum_{\bf k}g_{1j}({\bf k})e^{i{\bf k}{\bf l}}, \ j=1,2; \ l=0,1,..., p.
\label{g1j-consistency}
\end{equation}
Their right hand side is determined by a left hand side $g_{1j}(l)$ itself via Eqs. (\ref{Sigma4x4})-(\ref{Sigma(l)}), (\ref{omega12})-(\ref{g12}). They constitute an exact system of the $2(1+p)$ consistency equations. Its finding means a solution to the Ising problem in the same sense as finding of a self-consistency equation in the mean-field theory means a solution to a phase transition problem. It is straightforward to analyze these explicit consistency equations by the well-developed in the mean-field theory analytical and numerical methods. This result provides a regular way for the exact calculation of the critical exponents and the critical functions for the 3D systems.

\subsection{5.4. The constrained correlation functions and the order parameter}

    In the Ising model the order parameter is equal to the spontaneous magnetization of a spin at a lattice site,
\begin{equation}
\bar{S}^{'z}_{\bf r} \equiv \langle S^{z}_{\bf r} \rangle_{\mathcal{H}_s} = \langle s- \hat{\bar{\beta}}'_{\bf r}\hat{\beta}^{'}_{\bf r} \rangle_{\mathcal{H}_s} = s- \langle \hat{\bar{\beta}}'_{\bf r}\hat{\beta}^{'}_{\bf r}\hat{\theta} \rangle/P_s .
\label{orderIsing}
\end{equation}
For the spin value $s=1/2$, it is reduced to a quantity
\begin{equation}
\bar{S}^{'z}_{\bf r} = 1/2 - \rho'_{n_{\bf r}=1} , \qquad \rho'_{n_{\bf r}=n} =\langle \delta_{\hat{n}_{\bf r},n}\hat{\theta} \rangle /P_s , 
\label{spinIsing}
\end{equation}
determined by the $\hat{\theta}$-cutoff, true probability $\rho'_{n_{\bf r}=1}$ of the spin boson at the site ${\bf r}$ to have one quantum of excitation.  Note that the coherent order parameter in Eqs. (\ref{spinorder}) and (\ref{spinorder'}) is absent in the Ising model.

    The definition of the true, constrained Green's function in Eq. (\ref{spinGreen'}) in the Ising model can be rewritten in the form of Eq. (\ref{spinG'}), $G^{'J_2}_{J_1}= -\langle \tilde{b}^{J_2}_{J_1}[\tilde{\theta}_{\tau_1} \tilde{\theta}_{\tau_2}] \rangle /P_s$, via the partial two-operator contraction, taken for the function $f=\tilde{\theta}_{\tau_1} \tilde{\theta}_{\tau_2}$, which is determined by the cutoff factor in Eq. (\ref{spintheta}). We find an exact solution of the recurrence Eq. (\ref{spin2-contraction}) for that operator contraction at equal times:
\begin{equation}   
\tilde{b}^{I_2}_{I_1} = [(g^{-1})_{I_1}^{I_2} (1-\delta_{{\bf r_1},{\bf r_2}}) \theta(-\tilde{n}_{{\bf r_1}}) \theta(-\tilde{n}_{{\bf r_2}})+
\delta_{I_1,I_2}\theta(-\tilde{n}_{{\bf r_1}}) ]f .
\label{b(theta)Ising}
\end{equation}

    Using this result, we find the true, constrained correlation functions $g_{I_1}^{'I_2}$ of the spin bosons in a lattice:
\begin{equation}    
g_{I_1}^{'I_2} = -(g^{-1})_{I_1}^{I_2} (1-\delta_{{\bf r_1},{\bf r_2}}) \rho'_{n_{\bf r_1}=0,n_{\bf r_2}=0} -\rho'_{n_{\bf r_1}=0} \delta_{I_1,I_2} ,
\label{g'Ising}
\end{equation} 
where $\quad \rho'_{n_{\bf r_1}=n_1,n_{\bf r_2}=n_2} =\langle \delta_{\hat{n}_{\bf r_1},n_1}\delta_{\hat{n}_{\bf r_2},n_2}\hat{\theta} \rangle /{P_s}.$

\noindent They are determined by the matrix $(g^{-1})_{I_1}^{I_2}$, which is inverse to the matrix $g_{I_1}^{I_2}$ of the unconstrained correlation functions, calculated in the previous Sect. 5.3. The inverse matrix $g^{-1}$ can be calculated by a technique of the Toeplitz matrices, known from the theory of the 2D Ising model.

   A detailed analysis of the obtained results for the true correlation functions and order parameter as well as the true single-site and two-sites, joined probability distributions $\rho'_{n_{\bf r}}$ and $\rho'_{n_{\bf r_1},n_{\bf r_2}}$ of the spin boson occupations, which are the $\hat{\theta}$-cutoff versions of the calculated in Sect. 5.2 distributions $\rho_{n_{\bf r}}$ and $\rho_{n_{\bf r_1},n_{\bf r_2}}$, will be given elsewhere.
   
   Here we just present a remarkable exact result for the characteristic function $\theta_N \equiv \theta_N (u_1,...,u_N)=\langle \exp (i\sum_{{\bf r}={\bf r_1},...,{\bf r_N}}u_{\bf r}\tilde{n}_{\bf r}) \rangle$ of the $N$-sites, joined uncutoff probability distribution of the spin boson occupations 
\begin{equation}
\rho_{\{ n_{\bf r}\}} \equiv \rho_{n_1,...,n_N}=\langle \prod_{\bf r} \delta_{\tilde{n}_{\bf r},n_{\bf r}} \rangle.
\label{rhoNdefinition}
\end{equation}
Generalizing method of Eq. (\ref{Theta2}) for the $N$ sites, we find
\begin{equation}
\theta_N(u_1,...,u_N) = \prod_{\bf r} \Big[ \frac{1}{z_{\bf r}-1} \Big] \frac{1}{\sqrt{\text{det}(g+Z)}},
\label{thetaN}
\end{equation}
where $g$ is the correlation $2N\times 2N$-matrix $g_{I_1}^{I_2}$, calculated in Sect. 5.3; $Z=\text{diag}\{ z_I/(z_I-1) \}$. The characteristic functions in Eqs. (\ref{ThetaFinal}) and (\ref{Theta2Final}) for the single-site and two-sites, joined distributions, calculated in Sect. 5.2, immediately follow from this general result if one sets all other $u_{\bf r}$, except $u_1$ or/and $u_2$, to be zero, so that all irrelevant factors in Eq. (\ref{thetaN}) become unity and only the determinants of the $2\times2$- or $4\times4$-matrices remain. The joined occupations probabilities for the $N$ sites can be calculated simply by a differentiation,
\begin{equation}
\rho_{\{ n_{\bf r}\}} = \prod_{{\bf r}} \Big( \frac{1}{n_{\bf r}!} \frac{\partial^{n_{\bf r}}}{\partial z_{\bf r}^{n_{\bf r}}} \Big) \theta_N \Big|_{\{z_{\bf r}=0\}}, z_{\bf r}=z_I=e^{iu_{\bf r}},
\label{rhoN}
\end{equation}
similar to the calculation in the end of Sect. 5.2. Actually, due to the $\hat{\theta}$-cutoff in Eq. (\ref{spintheta}), only the first two occupation numbers, $n_{\bf r}=0, 1$, are of interest. Note that the uncutoff distributions $\rho_{n}$, $\rho_{n_1,n_2}$ in Eqs. (\ref{rhoJacobi}), (\ref{rhoTheta2}) and $\rho_{\{ n_{\bf r}\}}$ contain, in fact, all effects of the constraints and spin interaction (including the constraint interaction), except only the $\hat{\theta}$-cutoff in Eq. (\ref{spintheta}) for the second step of the Hilbert space reduction in Eq. (\ref{spinreduction}), since they are calculated for the exact, constrained and $\hat{\theta}$-cutoff, Hamiltonian in Eq. (\ref{HIsing}). The cutoff distributions $\rho'_{n}$, $\rho'_{n_1,n_2}$, and $\rho'_{\{ n_{\bf r}\}}$ are simply their $\hat{\theta}$-cutoff versions. 

    A crucial point is that the derived consistency equations (\ref{g1j-consistency}) are the exact ones and, contrary to the mean-field theory equations, are valid both inside the entire critical region and outside it. They are equations for the auto- and cross-correlations of neighboring spin bosons and cannot be reduced to a simpler equation for the order parameter alone, as it was attempted in the original Landau approach. The latter is possible only approximately and only far outside the critical region into a well-ordered phase. A solution of the derived exact equations (\ref{g1j-consistency}) is more rich, than the mean-field one, and includes a fine structure and all details of the critical phenomena. At the wings of the critical region these consistency equations describe a critical behavior with the critical exponents, different from the ones, predicted by the mean-field theory. So, they yield the renormalization group results, but are based on the regular, quantum-field-theoretical analysis of Green's functions, perfected to an exact rigorous analysis. Moreover, the exact solution of the Ising model, given by Eq. (\ref{g1j-consistency}), describes both the mesoscopic and macroscopic systems and is valid not only for the asymptotics at the wings of the critical region, but also for the critical functions at a central part of the critical region.

\section{6. Discussion of the exact microscopic equations for the magnetic phase transitions in a critical region}

    The derived above two pairs of equations (\ref{spinbetaEq})-(\ref{spinGEq}) and (\ref{spin1-contraction})-(\ref{spin2-contraction}) (for the unconstrained, auxiliary coherent order parameter and Green's functions of spin excitations and for the basis partial operator contractions, respectively) together with a pair of equations (\ref{spinbeta'})-(\ref{spinG'}) for the true, constrained coherent order parameter and Green's functions of spin excitations and Eq. (\ref{spinself-energy}) for the total irreducible self-energy constitute the exact system of equations for the magnetic phase transitions in the critical region. It is crucial that they provide a fully regular, exact description of all critical phenomena and critical fluctuations, starting from the microscopic Hamiltonian of spin interaction and the first principles of statistical physics. They fully preserve a nonlinear, nonanalytical, critical structure of various statistical and thermodynamic quantities in the critical region, contrary to many other approximate or phenomenological models, which usually start from some unjustified, unreliable, ad hoc assumptions, hypotheses or simplifications in the Hamiltonian or in the description, inconsistent with an actual critical behavior. Moreover, the derived exact equations have a canonical, universal, symmetric and highly branched structure. All these features facilitate the appropriate controllable approximations, for example in the kernels of the integral equations, and finding the correct solutions for the critical region in a regular way, both for the mesoscopic and macroscopic systems. 

    For a discussion of the structure of these fundamental equations, we refer to Sect. 3.5, since it is very similar to the structure of the corresponding BEC equations. The main difference is a presence of many local constraint functions $\theta(2s-\hat{n}_{\bf r})$, instead of one global constraint $\theta(N-\hat{n})$ in the BEC problem, and, hence, many corresponding discrete variables $m_{J}$ in Eqs. (\ref{spin1-contraction})-(\ref{spin2-contraction}) for the basis partial operator contractions.

    An important feature of the magnetic phase transitions in a system of spins is that the equations (\ref{spin1-contraction})-(\ref{spin2-contraction}) for the basis partial operator contractions with a cutoff function $f \propto \theta (m_x-1-\tilde{n}_x)$ can be solved exactly by a direct recurrence $m_x \to m_x+1$. Indeed, one starts from a zero boundary value $\theta (m_x-1-\tilde{n}_x)=0$ at $m_x=0$ and ends at the required argument $m_x=2s+1$ in just $2s+1=2, 3,\dots$ steps, depending on a spin value $s= 1/2, 1,\dots$.

    Another important fact is that the derived equations get considerably simplified in the usually considered case of an interaction only between the neighboring spins in a homogeneous system.

    These facts facilitate an exact solution of Eq. (\ref{spinself-energy}) for the self-energy in terms of a few exact correlation parameters for the coordination number $p$ of neighboring spins. It yields a regular way for a solution of the critical phenomena problem. Namely that method is used for the exact solution of the long-standing 3D Ising problem, outlined in Sect. 5. The obtained exact system of $2(p+1)$ consistency equations for the auto- and cross-correlations (\ref{g1j-consistency}) is much more nontrivial, than the approximate mean-field self-consistency equation for the order parameter. That results in a different critical behavior of the solution near the critical point and, in particular, in a deviation of the actual critical exponents for the 2D and 3D Ising models from the ones, predicted by the mean-field theory. Specifically, the difference comes from a different behavior of the exact solution for the self-energies and eigen-energies near a singular point ${\bf k}=0$ of the Fourier sum or integral (in a thermodynamic limit) for the true correlation functions at a long range $r \to \infty$,
\begin{equation} 
g'_{1j}({\bf r})= \Big( \frac{a}{L} \Big)^d \sum_{\bf k}g'_{1j}({\bf k})e^{i{\bf k}{\bf r}} \approx \Big( \frac{a}{2\pi} \Big)^d \int g'_{1j}({\bf k})e^{i{\bf k}{\bf r}} d^d {\bf k}.
\label{corr-function}
\end{equation}
On this basis, one can compare the long-range asymptotics of the latter integral $\sim r^{2-d-\eta}$, that is, the values of the corresponding critical exponent $\eta$, within the present exact theory, Wilson's $\varepsilon$-expansion in the renormalization group theory, Landau mean-field theory, and exactly solvable models. The corresponding detailed analysis will be presented elsewhere. 

    The present regular theory can be compared against a few known solutions for the exactly solvable models \cite{Baxter1989}. The latter usually are unrealistic and demonstrate a degenerate behavior, barely related to a typical behavior of the actual physical systems. An example is the 2D Ising model for a square lattice of spins. Its solution was found by Onsager \cite{Onsager} in 1944 and by Zamolodchikov \cite{Zamolodchikov} in 1989 for the case of zero and non-zero external magnetic field, respectively. However, the most interesting point is that the present theory opens a regular way for a solution of the actual 3D problems (see Sect. 5), for example, for the systems with the Ising or Heisenberg Hamiltonians, which were never solved before.

    An interesting insight of the rigorous theory is that a fundamental part in the process of the spin correlations and magnetic phase transition is playing by the spin-boson creation and annihilation operators in Eq. (\ref{spinhatbeta'}), rather than the spin operators themselves. In fact, the mean values of the spin operators in Eqs. (\ref{spinvectorbeta})-(\ref{spin+-beta}), that is a spontaneous magnetization of the spin lattice, should be calculated via the averages $\bar{S}^{'z}_{\bf r} = s- \langle \hat{\beta}^{'\dagger}_{\bf r}\hat{\beta}^{'}_{\bf r}\hat{\theta} \rangle/P_s$ and $\bar{S}^{'+}_{\bf r}=\langle \sqrt{2s-\hat{n}_{\bf r}} \hat{\beta}_{\bf r}^{'} \hat{\theta} \rangle/P_s$. The latter as well as the spin correlations and other statistical and thermodynamic quantities can be found from the coherent order parameter $\bar{\beta}_{\bf r}$ and Green's functions of the spin bosons by means of the basis partial operator contractions similar to and based on the calculation of the true, constrained order parameter and Green's functions in Eqs. (\ref{spinbeta'})-(\ref{spinG'}). 

   Obviously, an exact reduction of the many-body spin system to a constrained system with a purely Bose statistics (i.e., with the pure Bose commutation relations for the creation and annihilation operators of the system's excitations) in Sect. 4.1 means that the ferromagnetic and antiferromagnetic phase transitions in the lattice of spins are intimately related to those constraints. Usually an unconstrained Bose system is not subject to a phase transition. An example is the BEC in a system of Bose particles, which occurs only due to the particle-number constraint $(\ref{N})$. In a system without that constraint, for example, in a system of photons in a black-box cavity with absorbing walls, the BEC phase transition does not exist.

    Thus, a fundamental symmetry, which control the dynamics and statistics of the many-body system of spins and which is generally wider than a symmetry, broken in a particular magnetic phase transition, is a local gauge symmetry of the spin bosons for each spin of the lattice. (Note that a similar, local rotational symmetry is the one broken in a nematic phase of the liquid crystals and in a superfluid $^3$He.) In virtue of the Noether's theorem, the corresponding integrals of motion are the constraints in Eq. (\ref{spin}), which determine, via the $\theta(2s-\hat{n}_{\bf r})$-cutoff factors, the nontrivial structure of the spin excitations in Eqs. (\ref{spinalpha'})-(\ref{spinhatbeta'}), the physical, constrained many-body Hilbert space $\mathcal{H}_s = \otimes_{\bf r} \mathcal{H}_{{\bf r}s}$ in Eq. (\ref{spinalpha'=beta'}), and the Hamiltonian in Eqs. (\ref{HxyHz})-(\ref{HIsing}). A spontaneous appearance of a macroscopic structure of magnetization in a ferromagnet or an antiferromagnet, that is a regular structure of the mean values of spins $\bar{{\bf S}}_{\bf r}$ in the lattice, usually breaks only a subgroup of the fundamental symmetry group and does not correspond to any additional constraints. The dynamics and statistics of the system is governed by those fundamental constraints, while one of the integrals of motion of the to-be-broken-symmetry subgroup could even coincide with an operator of an observable order parameter. A mean value of the latter could appear and grow in a course of a phase transition from zero to a macroscopic value due to restructuring of a quantum-statistical state of the system. This is what the magnetization does, for example, in the case of a homogeneous ferromagnetic phase transition. Hence, a magnetization, despite being a conserved quantity, may vary in mean value with changing temperature or other control parameters.

    This is in contrast to the fundamental integrals of motion, which are the fixed c-number constraints on the total occupations (\ref{spin}) of two spin bosons at each site of the lattice and always remain equal to twice the spin value, $2s$. In the case of BEC in a gas (Sect. 3) there is only one such fundamental c-number constraint: The total occupation (\ref{N}) of all single-particle energy levels in a trap is fixed to be equal to the total number of particles $N$, loaded into a trap.

    Finally, the presented rigorous theory clearly proves that a used by many authors \cite {Dyson1956,Dembinski1964,Tyablikov1967} approximation of the nonanalytical functions $\theta(2s-\hat{n}_{\bf r})$ and $\sqrt{2s-\hat{n}_{\bf r}}$ in the interaction Hamiltonian (\ref{HxyHz}) by the c-numbers or the polynomials is valid only asymptotically for the parameters far outside the critical region into a well-ordered phase. It leads there to the Landau mean-field theory or the Dyson's spin waves theory and cannot correctly describe the critical fluctuations and a transition from a disordered phase to an ordered phase through a critical point.

\section{7. Summary}

    It is straightforward to extend the general method and structure of the microscopic theory of the critical phenomena, presented in \cite{PLA2015} and outlined in detail here, to other phase transitions in various fields of physics, including the physics of condensed matter and quantum fields. The purpose of this paper is to bring an attention to the remarkable opportunity for a regular solution of the problem of phase transitions across a critical point, which is common to so many physical systems.
    
    The universal form of the microscopic theory contains

   (a) a system of two equations for the unconstrained (auxiliary) order parameter and Green's functions
\begin{equation}
\bar{\beta}_{jx}= \check{G}^{(0)}[\check{\Sigma}[\bar{\beta}_{jx}]],
\label{UbetaEq}
\end{equation}
\begin{equation}
G_{j_1x_1}^{j_2x_2}+ \bar{\beta}_{j_1x_1}\bar{\beta}^{*}_{j_2x_2} = G_{j_1x_1}^{(0)j_2x_2} + \check{G}^{(0)}[\check{\Sigma}[G_{j_1x_1}^{j_2x_2}]], 
\label{UGEq}
\end{equation}

   (b) the equation for the total irreducible self-energy $\Sigma$, that determines a kernel in the first two equations (a),
\begin{equation}
\langle T_{\tau} [\tilde{\beta}_{j_1x_1}, \tilde{H}^{'}_{\tau_1}]\Delta\tilde{\bar{\beta}}_{j_2x_2} \rangle = (-1)^{j_1} \int_0^{1/T} \Sigma_{j_1x_1}^{jx}G_{jx}^{j_2x_2} d\tau, 
\label{Uself-energy}
\end{equation}

   (c) the two equations for the basis partial one-operator (Eqs. (\ref{s}) and (\ref{spins})) and two-operator (Eqs. (\ref{b}) and (\ref{spinb})) contractions
\begin{equation}
\tilde{s}_{J}[f(\{ m_{J'} \})]= \bar{\beta}_J f(\{ m_{J'} \})+ g_J^{J'} \Delta_{m_{J'}} \tilde{s}_{J'}[f(\{ m_{J'} \})],
\label{U1-contraction}
\end{equation}
$$\tilde{b}^{J_2}_{J_1}[f(\{ m_{J'} \})] = -g_{J_1}^{J_2}f(\{ m_{J'} \}) - g_{J_1}^{J'} g_{J'}^{J_2} \Delta_{m_{J'}}f(\{ m_{J'} \})$$
\begin{equation}
+g_{J_1}^{J'_1} g_{J'_2}^{J_2} \Delta_{m_{J'_1}} \Delta_{m_{J'_2}} \tilde{b}^{J'_2}_{J'_1}[f(\{ m_{J'} \})],
\label{U2-contraction}
\end{equation}

   (d) the following explicit formulas for the calculation of the true, constrained by the $\hat{\theta}$-cutoff, order parameter and Green's functions from the unconstrained ones in (a) and from the basis partial operator contractions in (c): 
\begin{equation}
\bar{\beta}^{'}_{J}= \langle \tilde{s}_J[\tilde{\theta}_{\tau_i}] \rangle /P , \quad P=\langle \hat{\theta} \rangle ,
\label{Ubeta'}
\end{equation}
\begin{equation}
G^{'J_2}_{J_1}= -\langle \{ \tilde{s}_{\bar{J}_2}[ \tilde{s}_{J_1}[\tilde{\theta}_{\tau_1}\tilde{\theta}_{\tau_2} ]] + \tilde{b}^{J_2}_{J_1}[\tilde{\theta}_{\tau_1}\tilde{\theta}_{\tau_2}] \} \rangle /P .
\label{UG'}
\end{equation}
The latter are similar to Eqs. (\ref{beta'}), (\ref{G'}) and (\ref{spinbeta'}), (\ref{spinG'}). All other statistical and thermodynamic quantities can be found from those unconstrained and true order parameters and Green's functions in a similar way.

    All these fundamental equations are written here in a short symbolic form that includes the particular cases of BEC (Eqs. (\ref{betaEq}), (\ref{GEq}) and (\ref{1-contraction}), (\ref{2-contraction})) and ferromagnetism (Eqs. (\ref{spinbetaEq}), (\ref{spinGEq}) and (\ref{spin1-contraction}), (\ref{spin2-contraction})). Formally, being considered for a given total irreducible self-energy $\Sigma$ separately from other equations, each of the Eqs. (\ref{UbetaEq}), (\ref{UGEq}), (\ref{U1-contraction}), and (\ref{U2-contraction}) is a linear equation. However, together with Eq. (\ref{Uself-energy}) for the self-energy $\Sigma$ they form, in fact, an essentially nonlinear system of equations since the self-energy $\Sigma$ (see Eqs. (\ref{self-energy}), (\ref{spinself-energy})) is determined by their solutions and Eqs. (\ref{UGEq}), (\ref{U1-contraction}), (\ref{U2-contraction}) include the solutions of Eqs. (\ref{UbetaEq})-(\ref{UGEq}) as coefficients. Remarkably, these equations are valid in the entire critical region even for a mesoscopic system. This result was achieved by formulating a theory at a more fundamental, operator level via an introduction of the recurrence Eqs. (\ref{U1-contraction})-(\ref{U2-contraction}) for partial operator contractions, which reproduce themselves under partial contraction operation.

    Various solutions of these exact equations and their application for various systems and models with the particular microscopic Hamiltonians will be discussed elsewhere. In Sect. 5 we outline them for a generic example of the exact solution to a long-standing 3D Ising problem.
     
    The microscopic theory of phase transitions in the critical region and mesoscopic effects are directly related to the numerous modern experiments on and numerical studies of the BEC of a trapped gas (including BEC on a chip), where $N\sim 10^{2}-10^{7}$, (see, for example, \cite{BrownTwissOnBECThresholdNaturePhys2012,Hadzibabic2013NaturePhys,Hadzibabic2013,EsslingerCriticalBECScience2007,TwinAtomBeamNaturePhys2011,Dalibard2012NaturePhys,Dalibard2010,RaizenTrapControl,Blume,BECinterferometerOnChip2013,EntanglementOnChipNature2010,chipBECNature2001}), superfluidity of $^4$He in nanodroplets ($N\sim 10^{8}-10^{11}$) \cite{HeDroplets} and porous glasses \cite{Reppy} as well as various magnetic phase transitions \cite{PatPokr,Fisher1986,CritPhen-RG1992,Goldenfeld,Kadanoff,Vicari2002}, magnetic nanoparticles for information storage applications \cite{MagnNanoparticleStorage2011}, metamagnetism \cite{MetamagnetismPRB2013}, magnetic phase coexistence \cite{Ferromagner+ChargePRL2007}, interplay between magnetic phase transitions and BEC in the spinor Bose gases \cite{SpinorBoseGasMagnetism2013}, nanomagnets \cite{Nanomagnets2014}, and so on. In particular, note that the rigorous analysis in Sect. 4.1 also includes a Matsubara-Matsuda mapping of 1/2 spins onto a lattice of bosons \cite{MatsubaraMatsuda1956} as a particular approximate version of the Holstein-Primakoff representation. It means that the presented microscopic theory of critical phenomena can be immediately applied also to the intensively studied, in the last two decades, BEC in quantum XY-ordered magnets, since that BEC is intimately related to the Matsubara-Matsuda mapping (see a recent review \cite{{BECinXYmagnets}}). 

   In conclusion, we present a rigorous microscopic theory of phase transitions across a critical point. It includes the exact equations for the coherent order parameter and Green's functions as well as the exact recurrence equations for the basis partial operator contractions. The theory rigorously accounts for the local and global constraints of the many-body Hilbert space, including the ones related to the spontaneously broken symmetry via the Noether's theorem. It also accounts for the nonpolynomial functions in the Hamiltonian, including the ones originated from the constraints of the Hilbert space (the constraint nonlinear interaction). At last, using this theory, we outline the exact solution for the long-standing 3D Ising model. 
   
   We especially emphasize the crucial role of the novel recurrence equations for the basis partial operator contractions, Eqs. (\ref{1-contraction})-(\ref{2-contraction}), (\ref{spin1-contraction})-(\ref{spin2-contraction}), (\ref{U1-contraction})-(\ref{U2-contraction}), which offer universal, powerful, and very convenient tools for the calculation of various quantities, describing the critical phenomena. As is shown, in the case of the 3D Ising model in Sect. 5, this method is truly universal: One has just to select an appropriate, for a given quantity in question, function $f$ and solve these recurrence equations by the regular methods. 

    In particular, in the present paper we touch upon the following capabilities of the rigorous microscopic theory: 

   (i) to provide a critical-region extension of the Gross-Pitaevskii and Beliaev-Popov equations for the BEC,

   (ii) to get a critical-region extension of the Dyson's spin waves theory,  

   (iii) to solve for the constraints and to reveal an exact Hamiltonian with the constraint interaction for the Bose-Einstein condensation in a mesoscopic system, 

   (iv) to establish a regular, based on Green's functions, method for a calculation of the critical functions and critical indexes for the mesoscopic and macroscopic systems,
   
   (v) to uncover the new classes of the exactly solvable models of phase transitions, including the most interesting three-dimensional ones, and to get solutions for them,

   (vi) to relate the phase transitions to the fundamental constraints, originated from the fundamental symmetry of the many-body system in virtue of the Noether's theorem, and, finally,

   (vii) to set a universal microscopic description of the spontaneous symmetry breaking in the critical region. 

\noindent It is truly remarkable that, as is shown in the paper, the exact microscopic equations for such different phase transitions, as a BEC one in an interacting gas and a ferromagnetic one in a lattice of spins, are completely similar.


\begin{thebibliography}{999}
\bibitem{Landau1937} L.D. Landau, Phys. Z. Sowjetunion \textbf{11}, 26 (1937) (Translated in JETP \textbf{7}, 19 (1937)).

\bibitem{LLV} L.D. Landau and E.M. Lifshitz, \textit{Statistical Physics, Part 1} (Pergamon, Oxford, 1981). 

\bibitem{LL} E.M. Lifshitz and L.P. Pitaevskii, \textit{Statistical Physics, Part 2} (Pergamon, Oxford, 1981).

\bibitem{Onsager} L. Onsager, Phys. Rev. \textbf{65}, 117 (1944).

\bibitem{Yang} C.N. Yang, Phys. Rev. \textbf{85}, 808 (1952).

\bibitem{Ginzburg1960} V.L. Ginzburg, Sov. Phys. Solid State \textbf{21}, 1824 (1960).

\bibitem{Levanyuk1959} A.P. Levanyuk, Sov. Phys. JETP \textbf{9}, 571 (1959).

\bibitem{AGD} A.A. Abrikosov, L.P. Gorkov, and I.E. Dzyaloshinskii, \textit{Methods of Quantum Field Theory in Statistical Physics} (Prentice-Hall, Englewood Cliffs, N.J., 1963).

\bibitem{HohenbergMartin} P.C. Hohenberg, P.C. Martin, Ann. Phys. \textbf{34}, 291 (1965).

\bibitem{Kondor1974} P. Szepfalusy, I. Kondor, Annals of Physics \textbf{82}, 1 (1974).

\bibitem{FetterWalecka} A.L. Fetter and J.D. Walecka, \textit{Quantum Theory of Many-Particle Systems} (McGraw-Hill, New York, 1971).

\bibitem{Anderson1984} P.W. Anderson, \textit{Basis Notions of Condensed Matter Physics} (Addison-Wesley, London, 1984).

\bibitem{Shi} H. Shi and A. Griffin, Phys. Rep. \textbf{304}, 1 (1998).

\bibitem{Shlyapnikov1998} P.O. Fedichev and G.V. Shlyapnikov, Phys. Rev. A \textbf{58}, 3146 (1998).

\bibitem{PitString} L. Pitaevskii and S. Stringari, \textit{Bose-Einstein Condensation} (Clarendon, Oxford, 2003).

\bibitem{RevModPhys2004} J.O. Andersen, Rev. Mod. Phys. \textbf{76}, 599 (2004).

\bibitem{ProukakisTutorial2008} N.P. Proukakis, B. Jackson, J. Phys.B \textbf{41}, 203002 (2008).

\bibitem{Wilson} K.G. Wilson and J. Kogut, Phys. Rep. \textbf{12}, 75 (1974).

\bibitem{PatPokr} A.Z. Patashinskii and V.L. Pokrovskii, \textit{Fluctuation Theory of Phase Transitions} (Nauka, Moscow, 1982).

\bibitem{Fisher1986} P.B. Weichman, M. Rasolt, M.E. Fisher, and M.J. Stephen, Phys. Rev. B \textbf{33}, 4632 (1986).

\bibitem{CritPhen-RG1992} J.J. Binney, N.J. Dowrick, A.J. Fisher, E.J. Newman, \textit{The theory of critical phenomena.  An introduction to the renormalization group} (Oxford Univ. Press, Oxford, New York, 1992).

\bibitem{Goldenfeld} N. Goldenfeld, \textit{Lectures on Phase Transitions and Renormalization Group} (Addison-Wesley, Reading, MA, 1992).

\bibitem{Kadanoff} L.P. Kadanoff, \textit{Statistical physics: statics, dynamics and renormalization} (World Scientific, Singapore, New Jersey, London, Hong Kong, 2000).

\bibitem{Cardy1996} J.L. Cardy, \textit{Scaling and Renormalization in Statistical Physics} (Cambridge Univ. Press, Cambridge, 1996).

\bibitem{Vicari2002} A. Pelissetto and E. Vicari, Phys. Rep. \textbf{368}, 549 (2002).

\bibitem{Berges2002} J. Berges, N. Tetradis, and C. Wetterich, Phys. Rep. \textbf{363}, 223 (2002).

\bibitem{Reichl} L.E. Reichl, E.D. Gust, Phys. Rev. A \textbf{88}, 053603 (2013).

\bibitem{Khinchin1943} A.Ya. Khinchin, \textit{Mathematical Foundations of Statistical Mechanics} (Dover, New York, 1949).

\bibitem{Ruelle1969} D. Ruelle, \textit{Statistical Mechanics: Rigorous Results} (Benjamin, New York, 1969).

\bibitem{Zubarev1971} D.N. Zubarev, \textit{Nonequilibrium statistical thermodynamics} (Consultants Bureau, New York, 1974).

\bibitem{Dobrushin1973} R.L. Dobrushin, B. Tirozzi, Commun. Math. Phys. \textbf{54}, 173 (1973).

\bibitem{Lebowitz1978} M. Aizenmann, S. Goldstein, J.L. Lebowitz, Commun. Math. Phys. \textbf{62}, 279 (1978).

\bibitem{Martin-Lof1979} A. Martin-L$\ddot{\text{o}}$f, \textit{Statistical Mechanics and the Foundations of Thermodynamics} (Springer, Berlin, 1979).

\bibitem{EllisJStatPhys2000} R.S. Ellis, K. Haven, B. Turkington, J. Stat. Phys. \textbf{101}, 999 (2000).

\bibitem{Zubarev1996} D. Zubarev, V. Morozov, G. R$\ddot{o}$pke, \textit{Statistical Mechanics of Nonequilibrium Processes, Volume 1 and 2} (Akademie Verlag, 1996).

\bibitem{PLA2015} V.V. Kocharovsky and Vl.V. Kocharovsky, Phys. Lett. A \textbf{379}, 466 (2015).

\bibitem{HolsteinPrimakoff} T. Holstein, H. Primakoff, Phys. Rev. \textbf{58}, 1098 (1940).

\bibitem{PRA2010} V.V. Kocharovsky and Vl.V. Kocharovsky, Phys. Rev. A \textbf{81}, 033615 (2010).

\bibitem{JPhys2010} V.V. Kocharovsky and Vl.V. Kocharovsky, J. Phys. A: Math. Theor. \textbf{43}, 225001 (2010).

\bibitem{PRA2014} S.V. Tarasov, V.V. Kocharovsky, and Vl.V. Kocharovsky, Phys. Rev. A \textbf{90}, 033605 (2014). 

\bibitem{JPhys2014} S.V. Tarasov, V.V. Kocharovsky, and Vl.V. Kocharovsky, J. Phys. A: Math. Theor. \textbf{47}, 415003 (2014).

\bibitem{KwokWoo} P.C. Kwok, W.F. Woo, Phys. Rev. A \textbf{3}, 437, 2039 (1971).

\bibitem{KochLasPhys2007} V.V. Kocharovsky and Vl.V. Kocharovsky, Laser Physics \textbf{17}, 700 (2007). 

\bibitem{KochJMO2007} V.V. Kocharovsky and Vl.V. Kocharovsky, J. Modern Optics \textbf{54}, 2491 (2007).

\bibitem{PDE-Cheng} S.S. Cheng, \textit{Partial Difference Equations} (Taylor \& Francis, London and New York, 2003).

\bibitem{DE-Agarwal} R.P. Agarwal, \textit{Difference Equations and Inequalities: Theory, Methods, and Applications, 2nd ed.} (Marcel Dekker, New York, 2000).

\bibitem{DE-Elaydi} S. Elaydi, \textit{An Introduction to Difference Equations, 3rd ed.} (Springer, New York, 2005).

\bibitem{PRA2000} V.V. Kocharovsky, Vl.V. Kocharovsky, M.O. Scully, Phys. Rev. A \textbf{61}, 053606 (2000).

\bibitem{Bogoliubov1947} N.N. Bogoliubov, Bull. Moscow State Univ. \textbf{7}, 43 (1947).

\bibitem{BogoliubovLectures1} N.N. Bogoliubov, \textit{Lectures on Quantum Statistics, Vol. 1: Quantum  Statistics} (Gordon and Breach Science Publishers, New York, London, Paris, 1967).

\bibitem{Zagrebnov} V.A. Zagrebnov, J.-B. Bru, Phys. Rep. \textbf{350}, 291 (2001).

\bibitem{Dyson1956} F.J. Dyson, Phys. Rev. \textbf{102}, 1217 (1956).

\bibitem{GA} M. Girardeau, R. Arnowitt, Phys. Rev. \textbf{113}, 755 (1959).

\bibitem{Girardeau1998} M.D. Girardeau, Phys. Rev. A \textbf{58}, 775 (1998).

\bibitem{BrownTwissOnBECThresholdNaturePhys2012} A. Perrin et al., Nature Phys. \textbf{8}, 195 (2012).

\bibitem{Hadzibabic2013NaturePhys} A.L. Gaunt, R.J. Fletcher, R.P. Smith, and Z. Hadzibabic, Nature Phys. \textbf{9}, 271 (2013).

\bibitem{Hadzibabic2013} A.L. Gaunt, T.F. Schmidutz, I. Gotlibovych, R.P. Smith, and Z. Hadzibabic, Phys. Rev. Lett. \textbf{110}, 200406 (2013).

\bibitem{EsslingerCriticalBECScience2007} T. Donner et al., Science \textbf{315}, 1556 (2007).

\bibitem{TwinAtomBeamNaturePhys2011} R. B$\ddot{u}$cker et al., Nature Phys. \textbf{7}, 608 (2011).

\bibitem{Dalibard2012NaturePhys} R. Desbuquois et al., Nature Phys. \textbf{8}, 645 (2012).

\bibitem{Dalibard2010} E. Mimoun, L. De Sarlo, D. Jacob, J. Dalibard, and F. Gerbier, Phys. Rev. A \textbf{81}, 023631 (2010).

\bibitem{RaizenTrapControl} M. Pons, A. del Campo, J.G. Muga, and M.G. Raizen, Phys. Rev. A \textbf{79}, 033629 (2009).

\bibitem{Blume} K. Nho, D. Blume, Phys. Rev. Lett. \textbf{95}, 193601 (2005).

\bibitem{BECinterferometerOnChip2013} T. Berrada et al., Nature Commun. \textbf{4}, 2077 (2013).

\bibitem{EntanglementOnChipNature2010} M.F. Riedel et al., Nature \textbf{464}, 1170 (2010). 

\bibitem{chipBECNature2001} W. H$\ddot{a}$nsel et al.,  Nature \textbf{413}, 498 (2001). 

\bibitem{HeDroplets} L.F. Gomez et al., Science \textbf{345}, 906 (2014).

\bibitem{Reppy} G.M. Zassenhaus and J.D. Reppy, Phys. Rev. Lett. \textbf{83}, 4800 (1999).

\bibitem{Schwinger1965} J. Schwinger, \textit{Quantum theory of angular momentum} (Academic Press, New York, 1965).

\bibitem{Sakurai} J.J. Sakurai, \textit{Modern Quantum Mechanics} (Addison-Wesley, Reading, Massachusetts, 1994).

\bibitem{Baxter1989} R.J. Baxter, \textit{Exactly solved models in statistical mechanics} (Academic Press, London, 1989).

\bibitem{Zamolodchikov} A.B. Zamolodchikov, Int. J. Mod. Phys. A \textbf{4}, 4235 (1989).

\bibitem{Dembinski1964} S.T. Dembinski, Physica \textbf{30}, 1217 (1964).

\bibitem{Tyablikov1967} S.V. Tayblikov, \textit{Methods in the quantum theory of magnetism} (Plenum Press, New York, 1967).

\bibitem{MagnNanoparticleStorage2011} N.A. Frey and S. Sun, \textit{"Magnetic nanoparticle for information storage applications", Ch. 3 in "Inorganic nanoparticles: synthesis, applications, and perspectives", ed. by C. Altavilla, E. Ciliberto} (CRC Press, Boca Raton, London, 2011).

\bibitem{MetamagnetismPRB2013} P.T. Barton, R. Seshadri, A. Llobet, and M.R. Suchomel, Phys. Rev. B \textbf{88}, 024403 (2013).

\bibitem{Ferromagner+ChargePRL2007} C. Sen, G. Alvarez, and E. Dagotto, Phys. Rev. Lett. \textbf{98}, 127202 (2007).

\bibitem{SpinorBoseGasMagnetism2013} D.M. Stamper-Kurn and M. Ueda, Rev. Mod. Phys. \textbf{85}, 1191 (2013).

\bibitem{Nanomagnets2014} T. Zhang, X.P. Wang, Q.F. Fang, and X.G. Li, Applied Physics Reviews \textbf{1}, 031302 (2014).

\bibitem{MatsubaraMatsuda1956} T. Matsubara and H. Matsuda, Prog. Theor. Phys. \textbf{16}, 569 (1956).

\bibitem{BECinXYmagnets} V. Zapf, M. Jaime, and C.D. Batista, Rev. Mod. Phys. \textbf{86}, 563 (2014).


\end{thebibliography}
\end{document}